\documentclass[prd,nofootinbib,preprint,superscriptaddress]{revtex4}
\pdfoutput=1

\ifx\pdfoutput\undefined
\usepackage[dvips,bookmarks=false]{hyperref}	
\else
\usepackage{hyperref}	
\fi
\hypersetup{colorlinks,bookmarksopen,bookmarksnumbered,citecolor=blue,
linkcolor=blue,pdfstartview=FitH,urlcolor=blue}

\usepackage{graphicx}
\usepackage{amsmath}
\usepackage{amssymb}

\newcommand{\co}{\mathcal{O}}

\newcommand{\be}{\begin{equation}}
\newcommand{\ee}{\end{equation}}
\newcommand{\beq}{\begin{equation}}
\newcommand{\eeq}{\end{equation}}

\newcommand{\vect}[1]{\boldsymbol{\rm #1}}

\makeatletter
\renewcommand{\fnum@table}{\textbf{\tablename~\thetable}}
\renewcommand{\fnum@figure}{\textbf{\figurename~\thefigure}}
\makeatother

\begin{document}
\pagestyle{plain}

\vspace*{1cm}
\preprint{FERMILAB-PUB-09-632-T}

\title{Global interpretation of direct Dark Matter searches after CDMS-II results\vspace*{1cm}}

\author{\textbf{Joachim Kopp}\vspace*{3mm}}
\email{jkopp AT fnal.gov}
\affiliation{Theoretical Physics Department, Fermi National Accelerator Laboratory, P.O. Box 500, Batavia, IL60510, USA}

\author{\textbf{Thomas Schwetz}\vspace*{3mm}}
\email{schwetz AT mpi-hd.mpg.de}
\affiliation{Max-Planck-Institute for Nuclear Physics,
             PO Box 103980, 69029 Heidelberg, Germany}

\author{\textbf{Jure Zupan}\vspace*{3mm}}
\email{jure.zupan AT cern.ch}
\affiliation{ Faculty of mathematics and physics, University of  Ljubljana, Jadranska 19, 1000 Ljubljana, Slovenia}
\affiliation{Josef Stefan Institute, Jamova 39, 1000 Ljubljana, Slovenia}


\begin{abstract}
\vspace*{5mm} We perform a global fit to data from Dark Matter (DM)
direct detection experiments, including the recent CDMS-II results. We
discuss possible interpretations of the DAMA annual modulation signal
in terms of spin-independent and spin-dependent DM--nucleus
interactions, both for elastic and inelastic scattering. We find that
for the spin-dependent inelastic scattering off protons a good fit to all
data is obtained. We present a simple toy model realizing such a scenario.
In all the remaining cases the DAMA allowed regions are disfavored by other
experiments or suffer from severe fine tuning of DM parameters with
respect to the galactic escape velocity. 
Finally, we also entertain the possibility that the two events observed in
CDMS-II are an actual signal of elastic DM scattering, and we compare the
resulting CDMS-II allowed regions to the exclusion limits from other
experiments.

In this arXiv version of the manuscript we also provide in appendix A
the updated fits including recent CoGeNT results.
\end{abstract}
\maketitle


\section{Introduction}

Direct detection of Dark Matter (DM) relies on signals due to energy
deposited from DM recoiling on matter in a detector. The DAMA
collaboration has provided strong evidence for an annually modulated
signal in the scintillation light from sodium iodine detectors. The
combined data from DAMA/NaI~\cite{Bernabei:2003za} (7 annual cycles)
and DAMA/LIBRA~\cite{Bernabei:2008yi} (4 annual cycles) with a total
exposure of 0.82~ton\,yrs shows a modulation signal with $8.2\,\sigma$
significance. The phase of this modulation agrees with the assumption
that the signal is due to the scattering of Weakly Interacting Massive
Particles (WIMPs) forming the DM halo of our Galaxy.

After a series of experiments with negative results, the CDMS
collaboration has recently presented an analysis where 2 events over
an expected background of $0.9\pm0.2$ events have been seen after
612~kg~days of exposure \cite{:2009zw}. The probability to observe two
or more background events is 23\%, which means that the two events
neither provide a statistically significant evidence for DM
interactions nor can they be rejected as a background. In this paper
we investigate both hypotheses and confront them with the results of
the other DM direct detection experiments.
Another recent direct detection result is due to XENON10
\cite{Angle:2007uj} that is searching for signs of DM scattering on
liquid xenon. In a recent re-analysis \cite{XENON:2009xb} of their
316.4 kg\,day exposure they found 13 events with expected background of
7.4 events, which was used to set limits on DM scattering cross
sections. 

Both of these very recent experimental results merit re-examination of
the DAMA signal and its consistency with the results from the other
direct detection experiments.  In the present manuscript we address
the following questions:
\begin{itemize}
\item
  Is it possible to reconcile the DAMA annual modulation signal with
  the constraints from all other experiments?  
\item
  If interpreted as DM signal, are the two events observed in CDMS
  consistent with results of other experiments?
\end{itemize}
We will focus on four classes of DM scattering cross sections that
cover a large set of DM models: the spin-dependent (SD) and
spin-independent (SI) WIMP--nucleon scattering that can be either
elastic or inelastic. We will use the shorthand notations eSD, eSI,
iSD, iSI, to denote these four classes.

Part of the answer to the first question is already well known from
the literature, since many of the interpretations of the DAMA signal
are in conflict with the constraints from other DM direct detection
experiments. For instance, recent analyses of the eSI case have been
performed in~\cite{Bottino:2007qg,Bottino:2008mf, Petriello:2008jj,
Feng:2008dz, Chang:2008xa, Fairbairn:2008gz, Savage:2008er,
Kim:2009ke,Feldstein:2009np}. We will show that this scenario gets
even more disfavoured due to the new CDMS and XENON10 results.
Similarly, the explanation of the DAMA signal due to the elastic
spin-dependent scattering (eSD)
is disfavoured by the strong constraints from
COUPP~\cite{Behnke:2008zza}, KIMS~\cite{Lee:2007qn}, and
PICASSO~\cite{Archambault:2009sm}. 

The case of inelastic spin-independent (iSI) scattering of a DM
particle to a nearly degenerate excited
state~\cite{TuckerSmith:2001hy} is also tightly constrained,
especially by CRESST-II~\cite{Angloher:2008jj} and
ZEPLIN-III~\cite{Cline:2009xd}, c.f.\ recent
analyses~\cite{Chang:2008gd,MarchRussell:2008dy, Cui:2009xq,
Arina:2009um, SchmidtHoberg:2009gn,Savage:2009mk}. However, as we will
show, the spin-dependent inelastic scattering (iSD) offers a viable
explanation of the DAMA signal, consistent with all other constraints.
To the best of our knowledge, this solution to DAMA has not been
discussed in the literature before.

Unlike DAMA the CDMS signal is statistically very weak. Even so,
focusing on the second question above, we also entertain the
hypothesis that the two events in CDMS are indeed due to DM and
perform a maximum likelihood fit allowing for the presence of a
signal. In the case of elastic scattering (eSI and eSD) we obtained a
very weak (at the $1\sigma$ level) indication for a positive DM
signal, and compare the corresponding allowed regions to the
constraints from all other experiments.

The outline of the paper is as follows. In section \ref{Event_rates}
we collect expressions for predicted event rates in DM scattering
experiments, in section \ref{Experiments} we describe details of
individual experiments relevant for the combined analysis, while in
section \ref{constraints} we then give the results of the fits. A
simple DM toy model leading to inelastic spin-dependent scattering is
presented in section \ref{simple_model}, followed by conclusions in
section \ref{conclusions}. In appendix we also include the interpretation 
of most recent CoGeNT result that appeared after the publication of the 
manuscript. 

In addition to the four general classes of DM models considered in
this work, there are also other proposals to explain the DAMA signal,
including for example mirror world DM~\cite{Foot:2008nw}, DM with
electric or magnetic dipole moments~\cite{Masso:2009mu}, leptophilic
DM \cite{Kopp:2009et}, resonant DM--nucleus
scattering~\cite{Bai:2009cd}, modified scattering due to DM form
factors \cite{Feldstein:2009tr, Chang:2009yt}, and atomic DM
\cite{Kaplan:2009de}.  The implications of the new CDMS results for
the above models is an interesting open question that, however, is
beyond the scope of this paper.

\section{Event rates}
\label{Event_rates}

The differential counting rate in a direct DM detection experiment (in units
of counts per energy per kg detector mass per day) is given by
\begin{align}
 \frac{dR}{dE_d} = \frac{\rho_0}{m_\chi} \frac{\eta}{\rho_\mathrm{det}} \,
   \int_{v>v_{\rm min}}\!d^3v \, \frac{d\sigma}{dE_d} \, v f_\odot(\vect{v}) \, ,
 \label{eq:dRdE-1}
\end{align}
where $E_d$ is the energy deposited in the detector,
$\rho_0$ is the local DM density (which we take to be $0.3\
\text{GeV}\,\text{cm}^{-3})$, $\eta$ is the number density of target
particles, $\rho_\mathrm{det}$ is the mass density of the detector and $d\sigma/dE_d$ is the
differential cross section for scattering on a target nucleus.  If
the target contains different elements (like in the case of the DAMA NaI
crystals), the sum over the corresponding counting rates is implied.

In eq.~\eqref{eq:dRdE-1}, $f_\odot(\vect{v})$ is the local WIMP velocity
distribution in the rest frame of the detector, normalized according
to $\int d^3v \, f_\odot(\vect{v}) = 1$. It follows from the DM
velocity distribution in the rest frame of the galaxy,
$f_\mathrm{gal}(\vect{v})$, by a Galilean transformation with the
velocity of the Sun in the galaxy and the motion of the Earth around
the Sun. For $f_\mathrm{gal}(\vect{v})$ we assume the conventional
Maxwellian distribution with $\bar v = 220\,\rm km\, s^{-1}$ and a
cut-off due to the escape velocity from the galaxy of $v_{\rm esc} =
650\,\rm km\, s^{-1}$: $f_\mathrm{gal}(\vect{v}) \propto
\exp(-\vect{v}^2/\bar{v}^2) - \exp(v_{\rm esc}^2/\bar{v}^2)$ for $v
\leq v_{\rm esc}$ and zero for $v > v_{\rm esc}$. The precise value of
the escape velocity has a negligible impact on our results for elastic
scattering, while in the inelastic case the precise shape of the tails
of the velocity distribution is important \cite{Kuhlen:2009vh}, see the
discussion in sec.~\ref{constraints}.

The lower limit of the integration in eq.~\eqref{eq:dRdE-1} is set by
the minimal velocity $v_{\rm min}$ that the incoming DM particle has
to have in order to be able to deposit an energy $E_d$ in the
detector. For the case of inelastic $\chi N\to \chi' N$ scattering it
is given by
\beq\label{vmin}
v_{\rm min}=\frac{1}{\sqrt{2m_N E_d}}\left(\frac{m_N
E_d}{\mu_{\chi N}}+\delta\right),
\eeq
where $\mu_{\chi N}=m_\chi m_N / (m_\chi+m_N)$ is the reduced mass of the
nucleus--DM system, with $m_N$ and $M_{\chi}$ the nucleus and DM masses respectively, while $\delta$ is the mass difference between
$\chi'$ and $\chi$. The same equation also applies to elastic $\chi
N\to \chi N$ scattering, with $\delta=0$. As observed in
ref.~\cite{TuckerSmith:2001hy} for appropriately chosen $\delta$ one
can suppress the signal in experiments where DM scatters on lighter
nuclei, while not significantly affecting the rate in DAMA (see also
\cite{Lisanti:2009am,Lisanti:2009vy}). Namely for $\delta\gg {m_N
E_d}/{\mu_{\chi N}}$ the minimal velocity $v_{\rm min}$ falls with
$m_N$. If the signal is coming from the tails of the velocity
distributions, the difference between lighter and heavier nuclei, such
as germanium vs.\ iodine, can be significant (for $v_{\rm min}> v_{\rm
esc}$ the scattering is completely absent). Furthermore, the
inelasticity also suppresses the low energy signal, changing the shape
of the expected event rate from an exponentially falling function of
the recoil energy to a bump-like signal at higher energies. This, in
addition, improves the fit to the DAMA modulated signal energy
spectrum.

The differential cross section for scattering on a target nucleus is
(per assumption) given by the spin independent (SI) and spin dependent
(SD) contributions, which are conventionally written as (see
e.g.\ \cite{Ullio:2000bv})
\beq
  \frac{d\sigma}{dE_d} = \frac{m_N}{2\mu_{\chi N}^2 v^2}\left(\sigma^{\rm SI} F^2(E_d)
                          +\sigma^{\rm SD} S(E_d)\right),
\eeq
where $\sigma^{\rm SI, SD}$ are the integrated SI and SD cross
sections for DM scattering on nucleus, but with form factors factored
out. For the SI form factor $F(E_d)$ we use~\cite{Jungman:1995df}
$F(E_d) = 3 e^{-\kappa^2 s^2/2} [\sin(\kappa r)-\kappa r\cos(\kappa
r)] / (\kappa r)^3$, with $s = 1$~fm, $r = \sqrt{R^2 - 5 s^2}$, $R =
1.2 A^{1/3}$~fm, $\kappa = \sqrt{2 m_N E_d}$ (and
$q^2\simeq-\kappa^2$). The SD form factor $S(E_d)$ is computed
according to ref.~\cite{Toivanen:2009zz} for $^{133}$Cs (abundant in
the CsI crystals used by the KIMS experiment) and according to
ref.~\cite{Bednyakov:2006ux} for all other nuclei.

Even though the form factors were factored out of the definitions of
$\sigma^{\rm SI, SD}$, these quantities still depend on nuclear structure through
isospin content (the number of protons vs.\ neutrons). The SI cross
section is thus
\beq \sigma^{\rm SI}=\frac{[Zf_p+(A-Z)f_n]^2}{f_p^2}\frac{\mu_{\chi
N}^2}{\mu_{\chi p}^2}\sigma_p^{\rm SI}, \eeq with A the atomic mass
number, Z the charge of the nucleus, $f_{p,n}$ the SI DM couplings to
proton and neutron respectivelly, $\mu_{\chi p}$ the reduced
DM--proton mass, and $\sigma_p^{SI}$ the SI cross section for
scattering of DM on a proton. In the fits we will assume $f_p=f_n$ for
definiteness and quote results in terms of $\sigma_p^{\rm SI}$. Since
the ratio $A/Z$ is similar for different nuclei this choice mostly affects
only the overall value of $\sigma_p^{SI}$, while it does not affect
the relative sizes of contributions from different experiments. It is
easy to rescale our results for different values of $f_p$ and $f_n$
through $\sigma_p^{SI}\to \sigma_p^{SI}/(Z/A+(1-Z/A)f_n/f_p)^2$.

The SD cross section depends in addition on the spin $J$ of the nucleus
\beq
\sigma^{\rm SD} S(E_d) = \frac{4 \mu_{\chi N}^2 \pi}{3 \mu_{\chi p}^2 a_p^2 (2 J + 1)}
                   [a_0^2 S_{00}(q) + a_0 a_1 S_{01}(q) + a_1^2 S_{11}(q)] \sigma_p^{\rm SD}\,,
\eeq
where $a_0 = a_p + a_n$ and $a_1 = a_p - a_n$ are combinations of the
DM couplings to protons $a_p$ and neutrons $a_n$, and $S_{00}(q)$,
$S_{01}(q)$, $S_{11}(q)$ are the spin-dependent nuclear structure
functions, which we compute according to \cite{Toivanen:2009zz} for $^{133}$Cs
and according to \cite{Bednyakov:2006ux} for all other nuclei.
In the fits we will assume $a_p=1,
a_n=0$, a choice that leads to a good fit of data. The reason is that
a spin-dependent interaction couples predominantly to un-paired
nucleons. Hence, for coupling only to protons DAMA can be compatible
with the remaining experiments since $^{127}$I has odd $Z$, but even
$N$ (i.e., one unpaired proton), while $^{73}$Ge, $^{129}$Xe,
$^{131}$Xe all have unpaired neutrons but even $Z$. For the other
extreme choice $a_p=0, a_n=1$ the DAMA signal is safely incompatible
with the other searches, see e.g.~\cite{Savage:2008er}. Simultaneous
couplings to proton and neutrons can still be allowed, depending on
the relative size of the two couplings.

\section{Description of experiments and analysis methods}
\label{Experiments}

In our fits we use the most sensitive experimental data sets available
to date, coming from the experiments DAMA/LIBRA, CDMS-II, XENON10,
ZEPLIN-III, CRESST-II, KIMS, and PICASSO. In this section we describe
each experiment in turn and comment on the sensitivity to each of the
four classes of DM interactions -- eSI, eSD, iSI, iSD.

\subsection{DAMA}
In 2008, the DAMA collaboration has published results of the combined
DAMA/NaI and DAMA/LIBRA
experiments~\cite{Bernabei:2003za,Bernabei:2008yi}, corresponding to
an exposure of 0.82~ton~yr for a target consisting of radiopure
NaI(Tl) crystals. They observe an annual modulation in the signal,
\begin{equation}\label{eq:DAMAsignal}
S(E,t) = S_0(E) + A(E)\cos\omega(t-t_0) \,,
\end{equation}
where $\omega = 2\pi/1$~yr, $t_0 = 152$~days.  In the fit we use the
signal region from 2 to 8~keVee of the spectrum from the combined
DAMA/NaI and DAMA/LIBRA (``DAMA'', for brevity) data given in fig.~9
of~\cite{Bernabei:2008yi}, divided into 12 bins.  
The data points above 8~keVee  are consistent with no modulation. Since no signal is predicted in that range from our DM models, they do not provide an additional constraint on the fit, and are thus ignored. Our analysis of the
DAMA data is analogous to the one presented
in~\cite{Fairbairn:2008gz,Kopp:2009et}.

The signal in DAMA is the energy deposited in scintillation light,
while the scattered nucleus is loosing energy both electromagnetically
and through nuclear interactions (phonon excitations).  This effect is
taken into account by the quenching factors that convert the total
nuclear recoil energy $E_d$ to the energy seen in the event by the
experiment, $q \times E_d$, with units of equivalent electron energy
(keVee). For DAMA $q_\mathrm{Na} = 0.3$ and $q_\mathrm{I} \simeq
0.09$~\cite{Bernabei:1996vj}. It is, however,
known~\cite{Drobyshevski:2007zj,Bernabei:2007hw} that some recoil
nuclei, namely those travelling along the crystal planes, will
\emph{not} suffer from quenching, but deposit essentially all their
energy electromagnetically (corresponding to $q = 1$). The fraction of
recoil nuclei for which this happens has been calculated in
\cite{Bernabei:2007hw}, but this calculation leaves room for some
debate~\cite{Graichen:2002kg,Gelmini:2009he,Feldstein:2009np}, since
the channeling effect has not been measured in the relevant energy
range so far. In the following we will discuss to what extent possible
explanations of the global data rely on the presence of the channeling
effect.

For the fit to the DAMA data we construct a $\chi^2$ function from the
annually modulated part of the signal
\begin{equation}\label{eq:chisq}
\chi^2_\mathrm{DAMA}(m_\chi, \sigma_p) = \sum_{i=1}^{12} \left(
\frac{A^\mathrm{pred}_i(m_\chi, \sigma_p) - A^\mathrm{obs}_i}{\sigma_i}
\right)^2 \,,
\end{equation}
where the sum is over the energy bins, and
$A^\mathrm{obs}_i$ ($\sigma_i$) are the experimental data points
(errors) in figure~9 of~\cite{Bernabei:2008yi}. 
We also impose a
constraint that the predicted unmodulated signal $S_0$ from DM scattering
should not exceed the one measured by DAMA (which consists of background and
signal) for any given energy bin. A recent simulation provides a background
estimate for DAMA \cite{Kudryavtsev:2009gd}. While this information could be
useful for future more refined fits, we take at this point a conservative
approach and let the background float freely in the fit.

We find the best fit point by minimising eq.~\eqref{eq:chisq} with
respect to WIMP parameters. Allowed regions in the $(m_\chi,
\sigma_p)$ plane for elastic scattering or in the $(m_\chi, \sigma_p,
\delta)$ space for inealstic scattering at a given CL are obtained by
looking for the contours $\chi^2(m_\chi, \sigma_p) =
\chi^2_\mathrm{min} + \Delta\chi^2({\rm CL})$, where
$\Delta\chi^2({\rm CL})$ is evaluated for the corresponding degrees of
freedom (dof), e.g., $\Delta\chi^2(90\%) = 4.6$ or
$\Delta\chi^2(99.73\%) = 11.8$ for 2~dof.

 \subsection{CDMS-II}
The most recent analysis of CDMS-II was performed on data taken
between July 2007 and September 2008 in four periods. Only Ge
detectors were used for the DM search with a total exposure of
612~kg~days. Two events were seen in the 10--100~keV energy window,
with recoil energies of 12.3~keVnr and 15.5~keVnr\footnote{Here, keVnr
refers to the actual \emph{nuclear} recoil energy, as opposed to the
equivalent electron energy reported by DAMA.}, while
$0.8\pm0.1\pm0.2$, $0.04^{+0.04}_{-0.03}$ and $0.03$--$0.06$
background events are expected from misidentified surface events,
cosmogenic background and neutron contamination, respectively. 
In our fits we take
into account that the signal efficiency drops from $32\%$ at $20$
keVnr to $25\%$ at both 10 keVnr and 100 keVnr by linear extrapolation.  We also
include the previous CDMS search with null result for exposure of
397.8~kg~days obtained between October 2006 and July 2007
\cite{Ahmed:2008eu}. We use a constant energy resolution of 0.2 keV
for the CDMS germanium detectors. 

In most part of our work we follow the CDMS collaboration and use the
data only to set an upper bound on a possible signal from DM. To this
aim we employ Yellin's maximum gap method~\cite{Yellin:2002xd}, which
by construction leads only to a bound (and never to a positive
signal), without any assumptions on the possible origin (background or
signal) of observed events. Only in sec.~\ref{sec:cdms-sig} we are
more speculative, and perform a maximum likelihood fit to the two
observed events assuming a model for the background (details of that
analysis are given in sec.~\ref{sec:cdms-sig}).
 
 \subsection{XENON10}
The XENON10 experiment (``XENON'', for short) searches for DM scattering on
xenon nuclei by measuring simultaneously the scintillation and ionization
signals in purified liquid xenon. Using a fiducial mass of 5.4 kg they
collected a data sample of 316.4 kg\,day between October 6, 2006 and February
14, 2007 \cite{Angle:2007uj}.  The latest analysis~\cite{XENON:2009xb} of this
data yields 13 events with an expected background of 7.4 in the 2.0--75.0~keVnr
window. This is a re-analysis of the same data used in \cite{Angle:2007uj},
where 10 events had been found in a smaller energy window from 4.5--26.9~keVnr.
There is still some controversy concerning the effective light yield in liquid
xenon, which is needed to translate the observed ionization signal into a
nuclear recoil energy~\cite{Sorensen:2008ec, Manzur:2009hp}. In our analysis,
we use the correction factors from~\cite{Sorensen:2008ec}, but we also discuss
the impact of using instead the data from~\cite{Manzur:2009hp}. For the
detection efficiency and background estimates, we use the numbers given in
table~I of \cite{XENON:2009xb}.  The energy resolution is computed according
to~\cite{Savage:2008er}
\begin{align}
  \frac{\sigma_E}{E} = \alpha \sqrt{\frac{\rm keV}{E}} + \beta
  \label{eq:Eres}
\end{align}
with $\alpha = 0.579$ and $\beta = 0.021$.
We analyze the XENON data using the maximum gap method~\cite{Yellin:2002xd}.

 \subsection{ZEPLIN-III}
Like XENON10, the ZEPLIN-III experiment is also a two-phase liquid
xenon time projection chamber experiment that has accumulated an
exposure of 847~kg\,days between February 27th and May 20th 2008
\cite{Lebedenko:2008gb}.  They observed 7 events in the 2--16 keVee
energy window (shown in fig.~16 of \cite{Lebedenko:2008gb}), which
corresponds to the nuclear recoil energy window of
10.7--30.2~keVnr. In our analysis we convert keVee into keVnr (and
vice-versa) using the scintillation light yields from fig.~15 of
\cite{Lebedenko:2008gb}. The detection efficiency is taken from
fig.~14 of \cite{Lebedenko:2008gb}, and the energy resolution is
assumed to be the same as in XENON10, eq.~\eqref{eq:Eres}. We analyze
the data by employing the maximum gap method.
  
\subsection{CRESST-II}
The commissioning run of CRESST-II~\cite{Angloher:2008jj} from March 27th to July 23rd, 2007
has a cumulative exposure of 47.9~kg~days on CaWO$_4$ crystals. Following \cite{Angloher:2008jj},
we use only tungsten recoils in our analysis.
There are 3 events in the 10--40~keVnr energy window at about 17~keVnr,
18~keVnr, and 33~keVnr.  We assume a constant efficiency of 90\% and a
constant energy resolution of 1~keV~\cite{SchmidtHoberg:2009gn}.  For
the statistical analysis, we again employ the maximum gap method.
 
 \subsection{KIMS}
The KIMS experiment has an exposure of $3409~{\rm kg\,days}$ taken with
low background CsI(Tl) crystals \cite{Lee:2007qn}.  Since no
information on the actual running periods was available to us, we
assume them to be equally distributed throughout the year.  We compute
the energy resolution according to eq.~\eqref{eq:Eres} with $\alpha =
0.582$ and $\beta = 0.0021$ \cite{Kim:2002proc} and assume the
detection efficiency to be 30\% at 3~keVee and 60\% at
5~keVee~\cite{Lee.:2007qn}.  Between these two energies, we use linear
interpolation. The quenching factor is computed from a fit to fig.~4
of \cite{Kim:2002proc}.  The KIMS data is reported in a window ranging
from 3~keVee to 11~keVee, divided into eight bins for each of the four
detector modules. 
We use
the $\Delta\chi^2$ method to analyze this data in the same way as the
unmodulated DAMA data.
To be conservative, we add penalties to
the $\chi^2$ only where the predicted event rate is larger than the
measured one, in order to make sure that only an upper bound is
obtained.

 \subsection{PICASSO}
The PICASSO experiment at SNOLAB~\cite{Archambault:2009sm} is very
different from the other experiments considered in this work because
it is based on the superheated droplet (bubble chamber) technique to
search for DM recoiling on ${}^{19}$F nuclei in a C$_4$F$_{10}$ target.
As we will see below, scattering on ${}^{19}$F is very sensitive to SD
interactions.  The experimental procedure is to measure the bubble
formation rate as a function of the temperature $T$. Since bubble
formation is possible only above a certain threshold energy $E_{\rm
thr}$ which depends on $T$ and on the particle species, a DM signal
would manifest itself as an increase in the bubble formation rate over
a certain $T$ interval.  We compute $E_{\rm thr}$ according to eq.~(2)
of \cite{Archambault:2009sm}.  The PICASSO data, corresponding to a
total exposure of $13.75 \pm 0.48$~kg~days, is analyzed separately for
the two best detectors (modules 71 and 72) with the $\Delta\chi^2$
method, where the background is assumed to have the
form $a [1 + \tanh \,b (T - T_0)] / 2.0$, and $a$, $b$, and $T_0$ are
free parameters determined by the fit. We also allow for a 14\%
uncertainty in the predicted signal~\cite{Archambault:2009sm}.

\section{Results}
\label{constraints}

In this section we present the results of our fits to the data. We
discuss elastic SI and SD scattering in sec.~\ref{sec:el}, and
inelastic SI and SD scattering in sec.~\ref{sec:in}. While in those
two subsections we follow the CDMS collaboration and use their latest data
only to set upper limits, in sec.~\ref{sec:cdms-sig} we become more
speculative and discuss a possible interpretation of the two observed
events as a signal from SI or SD elastic DM scattering.

\subsection{Elastic SI and SD scattering}
\label{sec:el}

\begin{figure}
  \begin{center}
    \includegraphics[width=\textwidth]{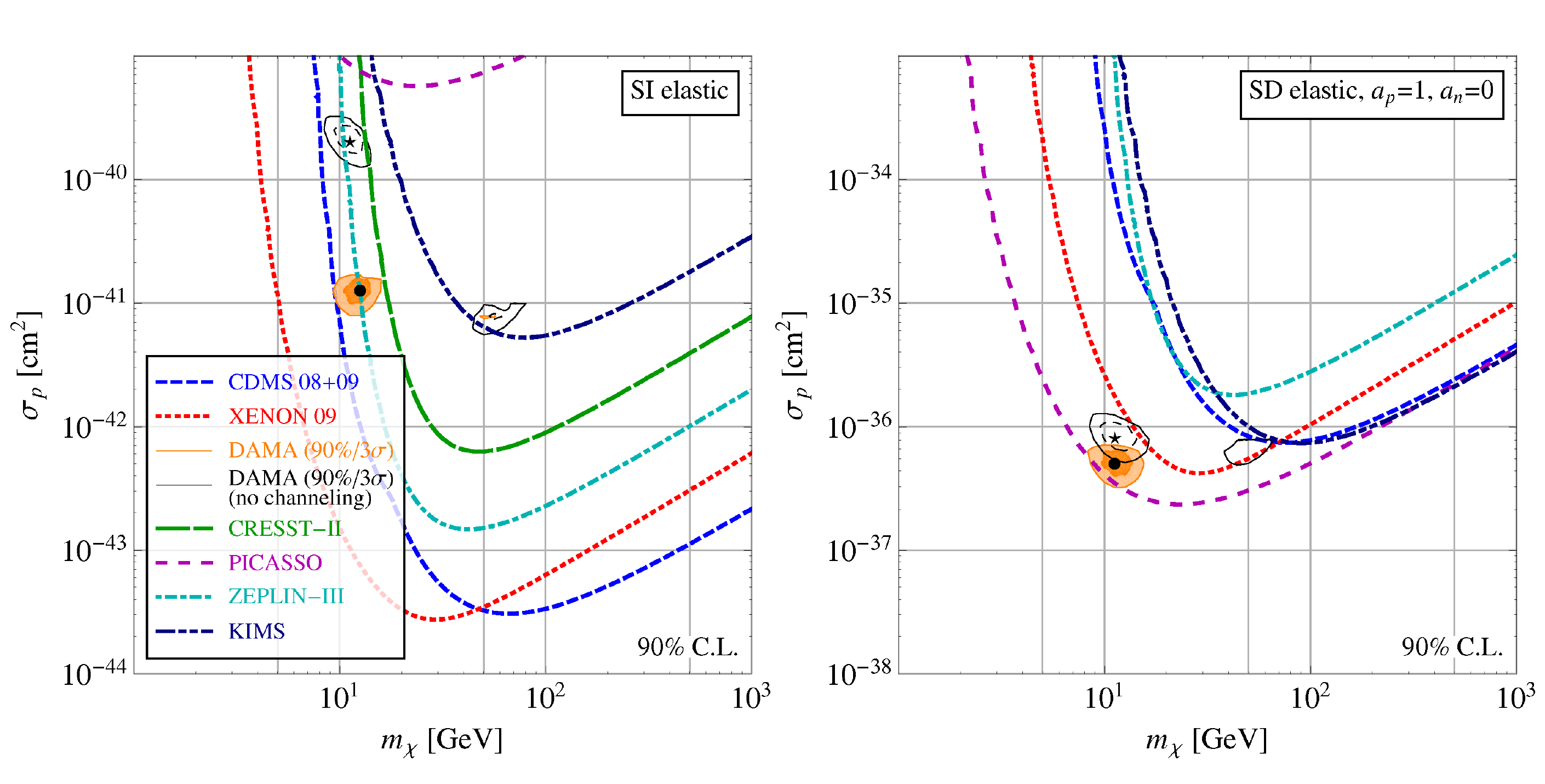}
  \end{center}
  \caption{DAMA allowed regions (90\% and 3$\sigma$~CL) and constraints
  from other experiments (90\%~CL) for SI scattering (left) and SD
  scattering off protons (right). Shaded DAMA regions have been
  obtained assuming the channeling effect according
  to~\cite{Bernabei:2007hw}, while the black contour curves correspond
  to no channeling.}
  \label{fig:msigma-plot-el}
\end{figure}

In fig.~\ref{fig:msigma-plot-el}, we summarize our results for elastic
scattering. We show the DAMA allowed regions compared to constraints
from other experiments in the plane of DM mass $m_\chi$ and the
interaction cross section for SI scattering and SD scattering off
protons. We observe that in both cases the DAMA regions are excluded
by the bounds. For the SI case (left panel) the most important bounds
come from CDMS and especially from XENON. They exclude the DAMA
regions regardless of the assumptions on channeling at high CL. Let us
note that for the DAMA region around 10~GeV without channeling
additional contraints from CDMS data on silicon apply (not
shown)~\cite{Fairbairn:2008gz}.

\begin{figure}
  \begin{center}
    \includegraphics[width=0.5\textwidth]{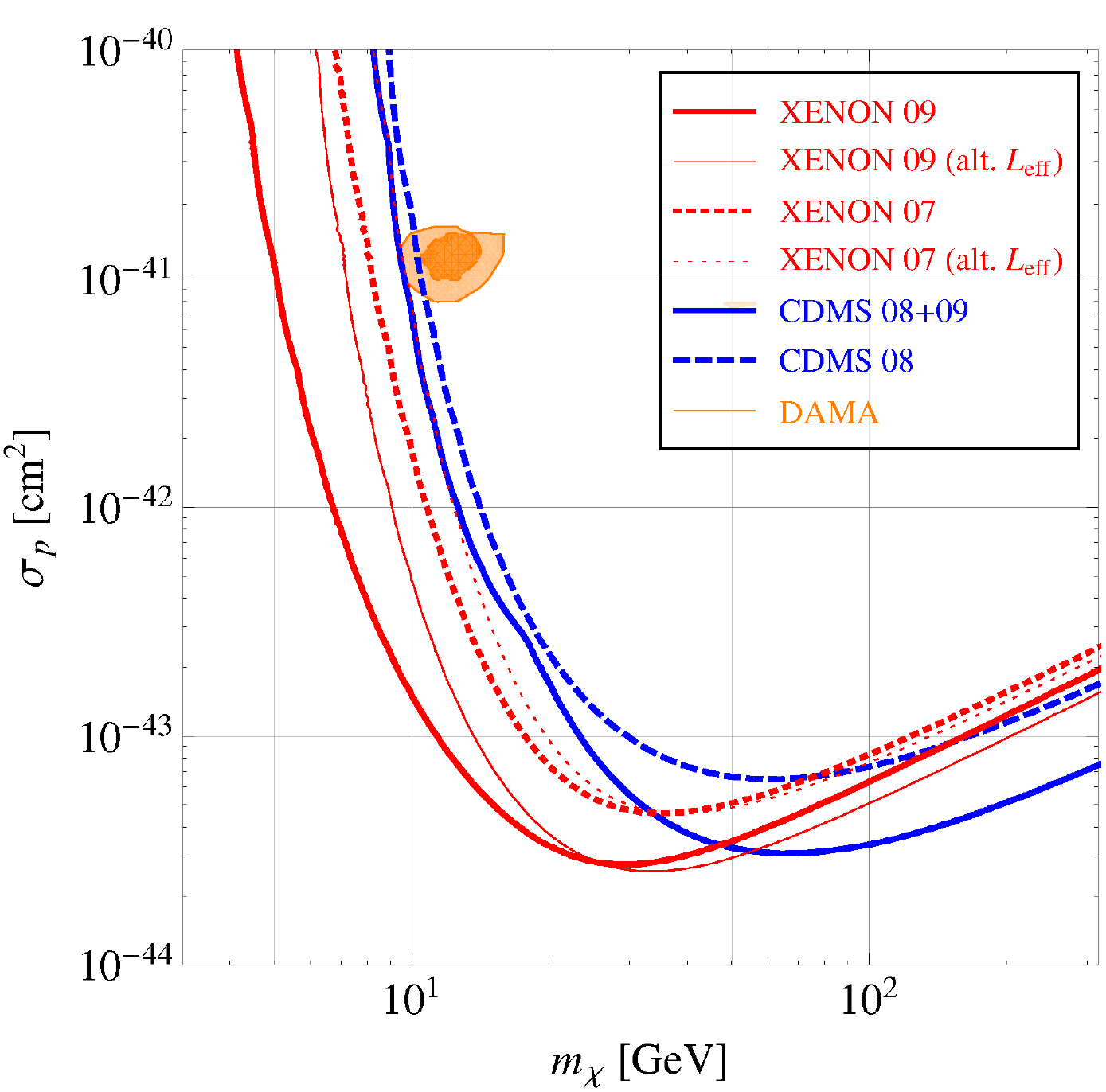}
  \end{center}
  \caption{Comparison of different data sets for CDMS and XENON for SI
    elastic scattering. XENON~07 refers to the analysis from
    \cite{Angle:2007uj}, while XENON~09 corresponds to the re-analysis
    of the same data from \cite{XENON:2009xb}. The CDMS 2008 and 2009
    data sets are from \cite{Ahmed:2008eu} and \cite{:2009zw},
    respectively. For XENON, different assumptions on the effective
    light yield $L_{\rm eff}$ are used. Thick curves are based on the
    measurement \cite{Sorensen:2008ec}, while thin curves are based on
    the alternative data set \cite{Manzur:2009hp}.}
  \label{fig:cdmsxenon}
\end{figure}

To study the impact of the new 2009 analyses from XENON \cite{XENON:2009xb} and
CDMS \cite{:2009zw}, we compare in fig.~\ref{fig:cdmsxenon} the old and new
data sets, and show also the impact of different assumptions on the effective
light yield $L_{\rm eff}$ in XENON.  We observe that despite the large
exposure, the new CDMS data has very small impact on the lower bound on
$m_\chi$. The reason is that the two observed events are located at low
energies, which are most relevant if $m_\chi$ is small, and therefore the
maximum gap method leads to a not so strong limit. The new CDMS data is more
important if $m_\chi$ is large so that a signal is expected also at larger
recoil energies, where no events have been observed. Therefore, thanks to the
large exposure the limit improves.
We also note a rather significant improvement in the low-mass limit
from XENON due to the 2009 analysis. There are two reasons for this
effect: first, the energy threshold has been lowered from
4.6~keVnr~\cite{Angle:2007uj} to 2~keVnr~\cite{XENON:2009xb}, and
second, the one event located close to 4.6~keVnr in the 2007 analysis
has been eliminated in the 2009 analysis. This leads to a large energy
interval at low recoil energies without events, which improves the
limit for low masses. 

Considering the case of SD elastic DM--nucleus scattering (right panel
of fig.~\ref{fig:msigma-plot-el}), we observe that rather strong
constraints come from PICASSO, if DM couples mainly to protons
($a_p = 1$, $a_n = 0$). Assuming channeling according
to~\cite{Bernabei:2007hw} the DAMA region at 90\%~CL is excluded by
the 90\%~CL bound from PICASSO, while both experiments are marginally
compatible at $3\sigma$. Without the channeling effect, there is no
overlap of allowed regions.
We do not show the case of
SD scattering off neutrons ($a_p = 0$, $a_n = 1$), since in that case
the DAMA region is safely excluded by CDMS and XENON
\cite{Savage:2008er}, see also fig.~\ref{fig:cdms-allowed}. The reason is
that the ${}^{19}$F nuclei in the PICASSO experiment have an unpaired
proton, while the spin-sensitive isotopes in CDMS (${}^{73}$Ge) and
XENON (${}^{129}$Xe, ${}^{131}$Xe) have an unpaired neutron. Let us
mention that in the proton case, also the COUPP~\cite{Behnke:2008zza}
experiment provides a relevant constraint. However, it was not
possible for us to implement a simulation of COUPP using the available
information.

\subsection{A signal in CDMS?}
\label{sec:cdms-sig}

Our default analysis of CDMS data uses the maximum gap
method~\cite{Yellin:2002xd}, which produces by construction only an
upper limit on a DM signal. Let us now be somewhat more speculative
and interpret the two events observed in the latest CDMS data as
positive signal from DM scattering (see also~\cite{Bottino:2009km}).
In order to do this we use a model for the energy shape of the
expected background from surface events. According to \cite{:2009zw},
this background is estimated by using the measured events in the
$2\sigma$ window from their previous analysis, where no timing cut is
yet imposed, see figure~3 of \cite{Ahmed:2008eu}.  We count the number
of events in the signal region and perform a fit to the energy
distribution.  Normalizing to the expected total number of background
events (i.e.\ 0.8), we find $dN^{\rm bkgr}/dE_d = -0.00295 +
0.463/E_d$ where the recoil energy $E_d$ is in keVnr. Using this
parameterization for the expected background shape we perform a fit to
the two events by using the so-called extended maximum likelihood
method \cite{Barlow:1990}.

\begin{figure}
  \begin{center}
      \includegraphics[width=\textwidth]{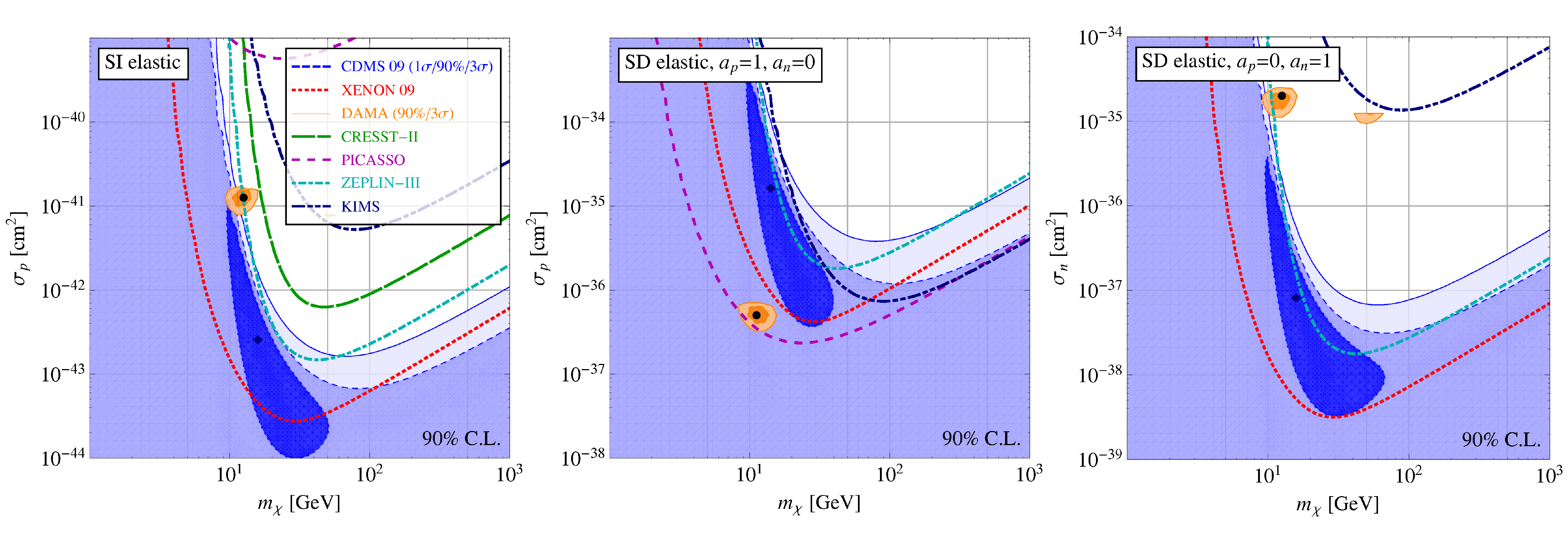}
  \end{center}
  \caption{Allowed regions for CDMS 2009 data (1$\sigma$, 90\% and
  3$\sigma$~CL), DAMA (90\% and 3$\sigma$~CL), and constraints from
  other experiments (90\%~CL) for elastic SI scattering (left), SD
  scattering off protons 
  (middle), and SD scattering off neutrons
  (right).}
  \label{fig:cdms-allowed}
\end{figure}

In fig.~\ref{fig:cdms-allowed} we show the CDMS allowed regions
compared to the constraints from the other experiments for SI and SD
elastic scattering. CDMS allowed regions are defined by $\Delta\chi^2
\equiv -2\log\mathcal{L/L^\mathrm{max}}$ contours for 2~dof. We find
that at 1$\sigma$~CL a closed allowed region appears for CDMS
(``positive signal''), while already at 90\%~CL only an upper bound is
obtained. The CDMS favoured region is largely excluded by the XENON
bound (and the PICASSO bound, in case of SD scattering off
protons). In the case of SI scattering we have checked that in the
combined fit to all all experiments except DAMA no closed region appears even at
1$\sigma$, and we obtain only an upper limit.

We have also verified that in the case of inelastic scattering no
closed region appears for CDMS. This follows from the fact that for
inelastic scattering the signal is shifted to larger recoil energies,
where CDMS sees no events, and therefore the data do not favour a
positive signal. Hence, in the following we return to the conservative
approach and use CDMS data only to set an upper limit on a DM signal
using the maximum gap method.

\subsection{Inelastic SI and SD scattering}
\label{sec:in}

Let us now consider the assumption of inelastic DM scattering, put
forward in ref.~\cite{TuckerSmith:2001hy} in order to reconsile the
DAMA signal with constraints from other experiments exploring the
modified kinematics, see eq.~\eqref{vmin} and the discussion given
there. We start by first presenting results for the SI case, largely
discussed in the literature, and then extend the inelastic
scattering hypothesis also to SD interactions. 

In order to identify possible solutions we have performed a combined
analysis of all experiments, by adding the $\chi^2$ functions of the
individual experiments. For those experiments which we analyse with
the help of the maximum gap method (see sec.~\ref{Experiments}) we
proceed as follows. For a given point in the parameter space the
probability obtained by the maximum gap method is converted into a
$\Delta\chi^2$ by inverting the integral over the
$\chi^2$-distribution for 2~dof. This ``fake'' $\Delta\chi^2$ is added
to the one from the remaining experiments. Then we perform a scan over
all three model parameters, $m_\chi$, $\sigma_p$, and $\delta$, in
order to search for local minima in the total $\chi^2$.  While this
recipe to incorporate the maximum gap method leads only to approximate
confidence regions, it suffices for our purpose to locate potential
solutions in the 3-dimensional parameter space. 
\begin{figure}
  \begin{center}
    \begin{tabular}{cc}
      \hspace{.25cm}\includegraphics[width=7.5cm]{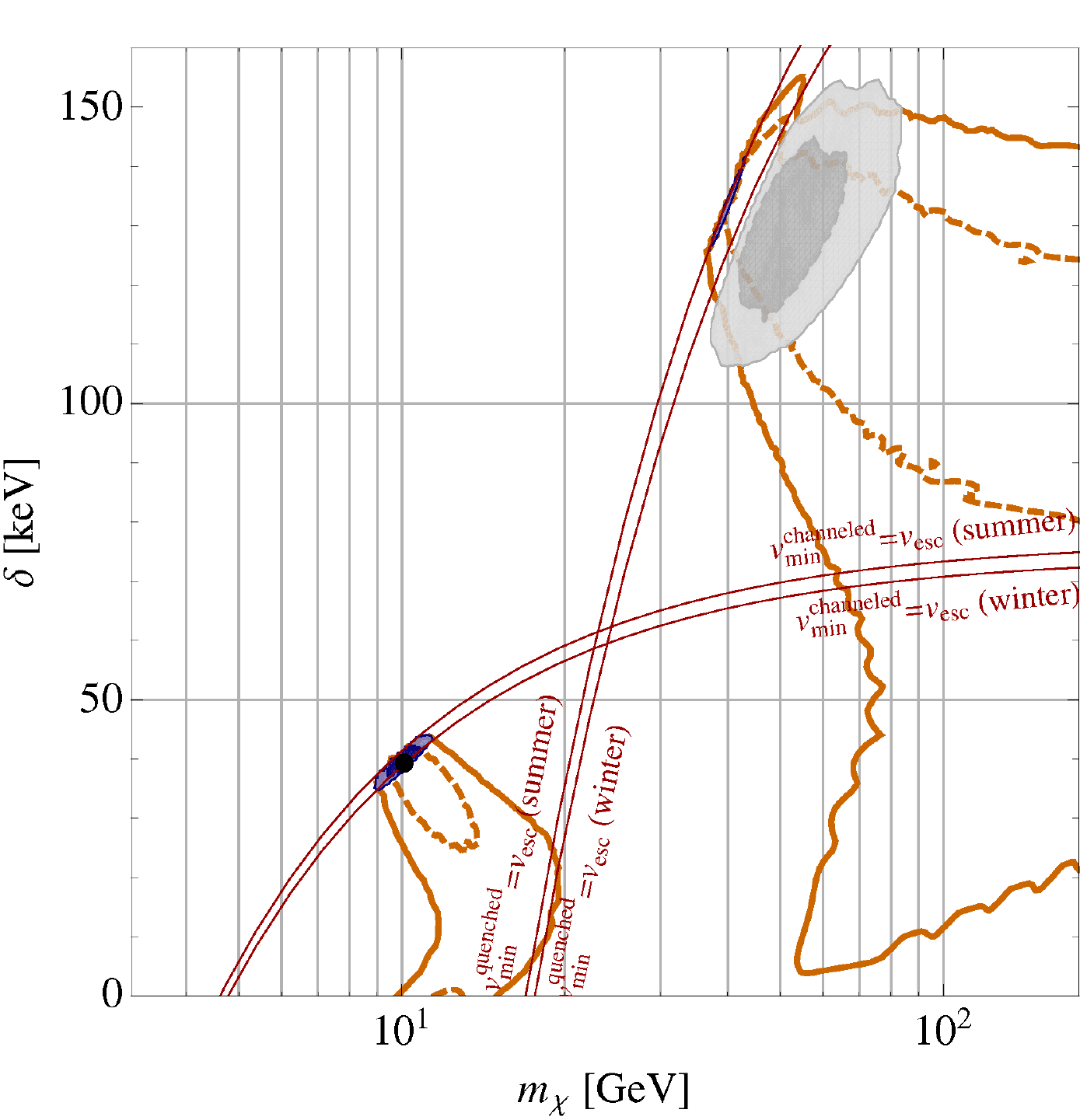} & \\
      \includegraphics[width=8cm]{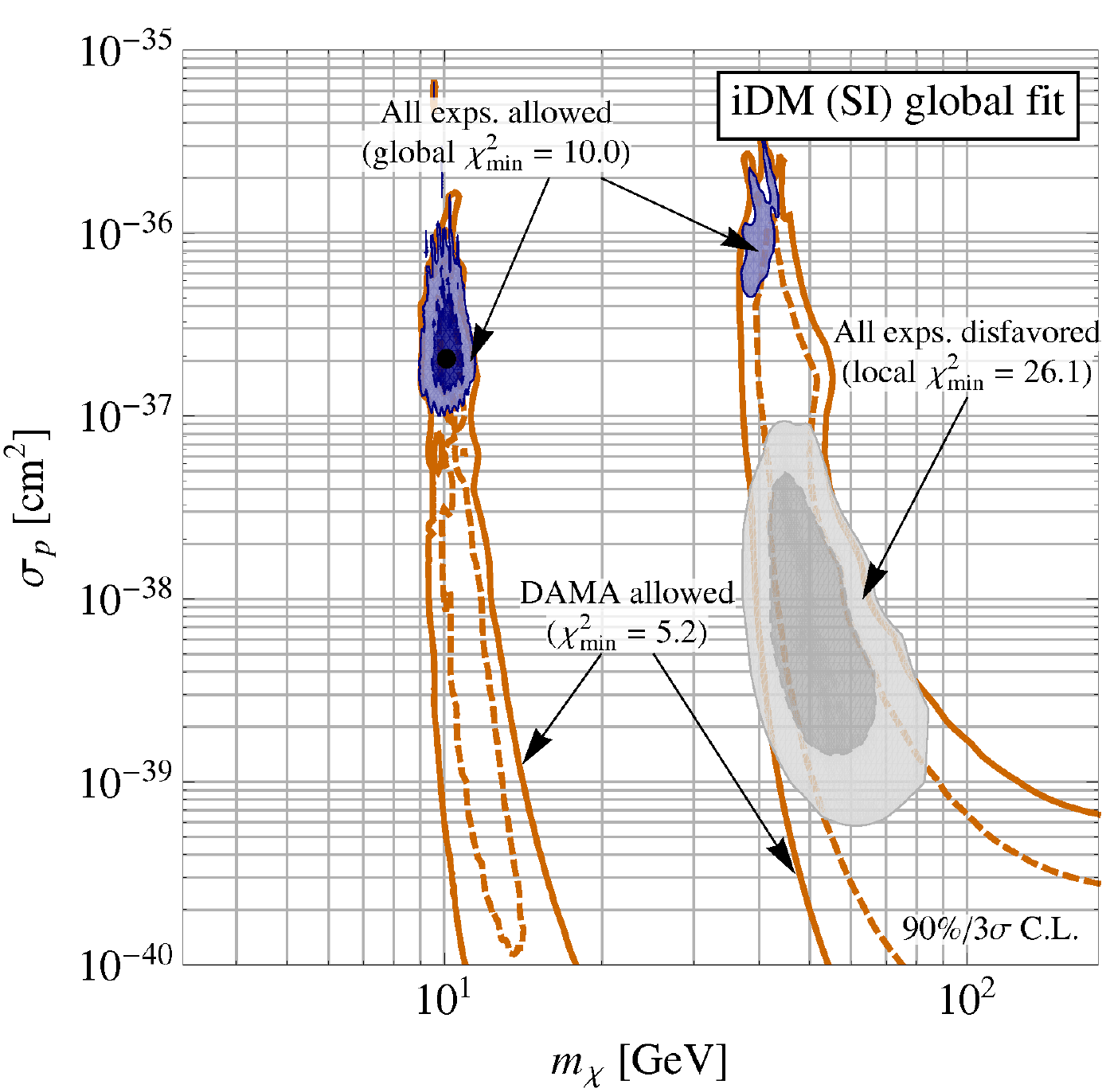} &
      \raisebox{.15cm}{\includegraphics[width=8cm]{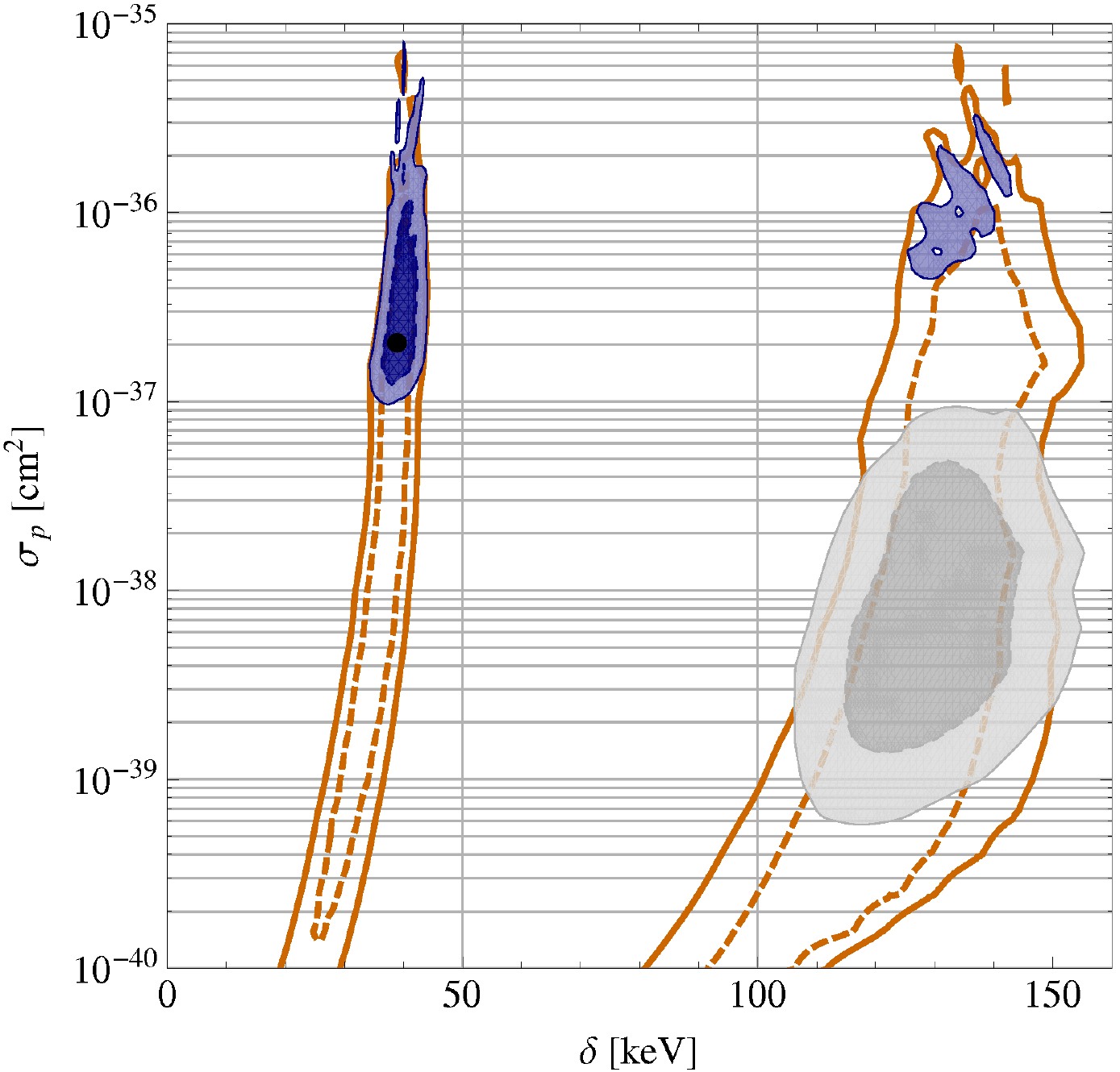}}
    \end{tabular}
  \end{center}
  \caption{Global fit of inelastic, spin-independent DM. We show
  projections of the 3-dimensional regions at 90\% and 3$\sigma$ CL
  onto the three 2-dimensional planes, by minimizing the global
  $\chi^2$ in each case with respect to the third (un-displayed)
  parameter. Confidence regions are defined for 2~dof. The fit
  includes CDMS (2008 + 2009 data), XENON (2009 analysis), DAMA,
  CRESST-II, ZEPLIN-III, and KIMS. The dark shaded regions are the allowed regions defined
  with respect to the global minimum, while the light shaded regions
  are defined relative to a local minimum, which by itself is
  disfavored relative to the global minimum with $\Delta\chi^2 =
  16.1$. The open contours correspond to DAMA data only. In the upper
  panel we show also contours of $v_\mathrm{min} = v_\mathrm{esc} +
  v_\mathrm{earth}$ for a recoil energy of $E = 3$~keV, with and
  without quenching of iodine scatters, where $v_\mathrm{earth}$ is
  the velocity of the earth relative to the halo, depending on the
  time in the year.}
  \label{fig:idm-fit}
\end{figure}

Fig.~\ref{fig:idm-fit} shows the results of such a scan for the SI
case.  We show projections of the 3-dimensional allowed regions onto
the three 2-dimensional planes. We identify three local minima. The
global minimum is $\chi^2_\mathrm{min, glob} = 10.0$ and is located at
\beq\label{eq:bfp}
\sigma_p = 2\times 10^{-37} \, {\rm cm^2}\,,\quad
m_\chi = 10.1\,{\rm GeV} \,,\quad
\delta = 39\, {\rm keV} \,.
\eeq
This solution corresponds to channeled events from
iodine~\cite{SchmidtHoberg:2009gn} and relies on the channeling
calculations from~\cite{Bernabei:2007hw} 
(scattering on sodium does not contribute to the signal, since the kinematics of inelastic scattering favor heavy over light nuclei). Furthermore this solution
corresponds actually to a tiny, rather fine tuned region in $\delta$
and $m_\chi$. The values are choosen such that the minimal DM velocity
$v_\mathrm{min}$ required to give a recoil energy of 3~keV needed to
explain the DAMA signal is very close  the galactic escape
velocity. Indeed, the parameters are tuned such that the signal in
DAMA is non-zero only in summer but zero in winter. This maximally
enhances the modulated signal, while at the same time suppresses the
unmodulated rate as well as the signal in the other
experiments.\footnote{We remark that in such a case the fit to the
DAMA modulated signal in terms of a cosine according to
eq.~\eqref{eq:DAMAsignal} might not provide a good description, since
the signal would correspond to a truncated cosine. We are not
exploring such additional signatures here.} This is illustrated in the
upper panel of fig.~\ref{fig:idm-fit}, where we show that the allowed
region is located precisely between the contour curves for
$v_\mathrm{min} = v_\mathrm{esc}$ in summer and in winter.  Because of
this fine tuning the best fit value for the cross section given in
eq.~\eqref{eq:bfp} is very sensitive to the precise implementation of
the DM halo profile and minor modifications in the analysis. This
tuning of DM parameters $\delta$ and $m_\chi$ relative to properties
of the galactic halo ($v_\mathrm{esc}$) is rather un-natural, and we
consider this solution as being disfavoured despite the formally very
good $\chi^2$ value. The same arguments apply for a local minimum
around $\sigma_p = 10^{-36} \, {\rm cm^2}, m_\chi = 40\,{\rm GeV},
\delta = 130\, {\rm keV}$, corresponding to quenched events on iodine.

\begin{figure}
  \begin{center}
    \includegraphics[width=\textwidth]{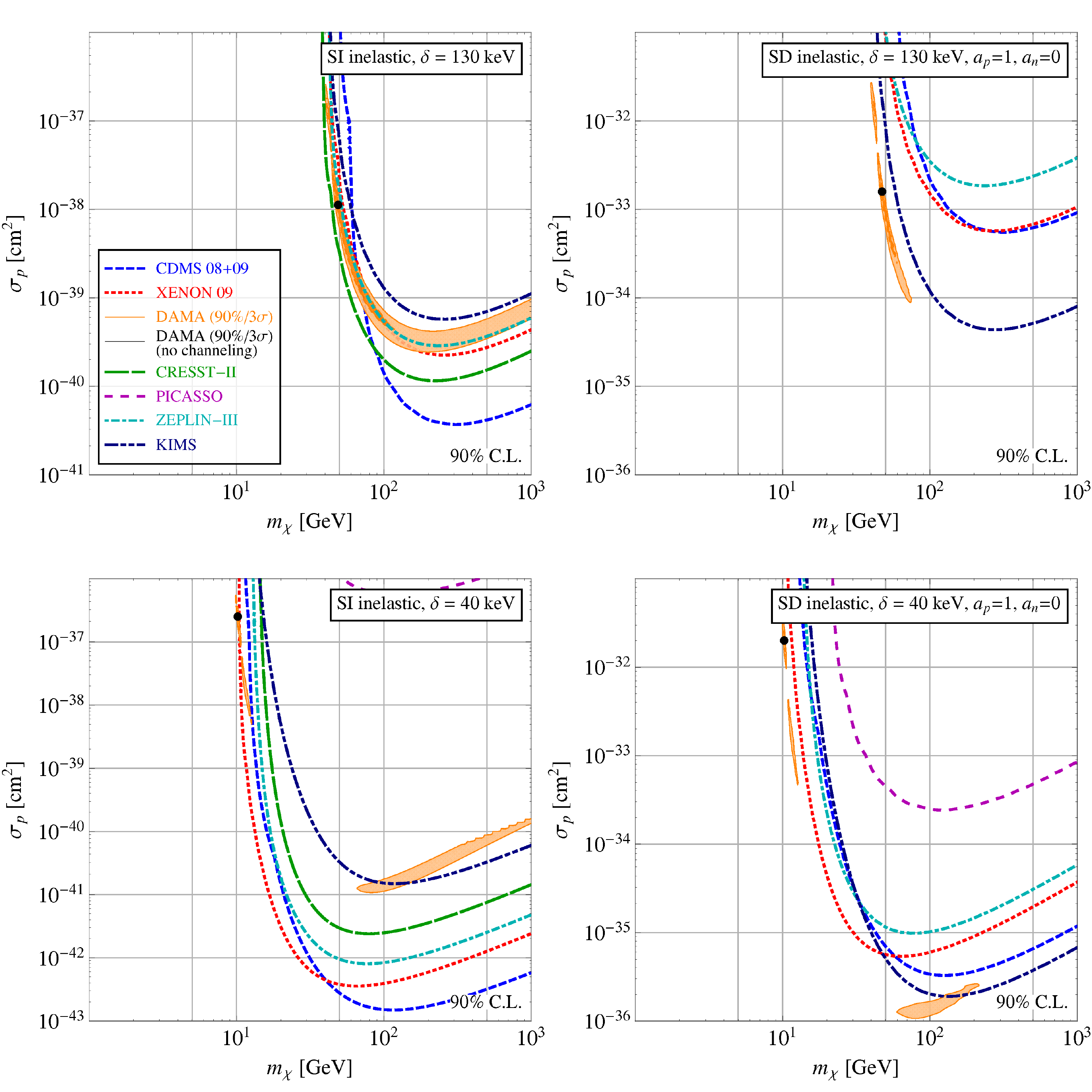}
  \end{center}
  \caption{DAMA allowed regions (90\% and 3$\sigma$~CL) and
  constraints from other experiments (90\%~CL) for inelastic DM
  scattering with SI interactions (left) and SD interactions with
  protons (right). We show regions in the $(m_\chi, \sigma_p)$ plane
  (2~dof) for fixed DM mass splitting $\delta$. The upper panels
  ($\delta = 130$~keV) correspond to a signal in DAMA from quenched
  events on iodine, whereas the lower panels ($\delta = 40$~keV)
  correspond to channeled events on iodine according
  to~\cite{Bernabei:2007hw}.}
  \label{fig:msigma-plot-inel}
\end{figure}

Another local minimum appears around $\sigma_p = 10^{-38} \, {\rm
cm^2}, m_\chi = 50\,{\rm GeV}, \delta = 130\, {\rm keV}$ (light shaded
regions in fig.~\ref{fig:idm-fit}). This corresponds to the
conventional iDM solution discussed recently by many authors,
e.g.~~\cite{Chang:2008gd,MarchRussell:2008dy, Cui:2009xq,
Arina:2009um, SchmidtHoberg:2009gn,Savage:2009mk}. This solution does
not suffer from the fine tuning problem, but we find that it is
disfavored with respect to the best fit with $\Delta\chi^2 =
16.1$. The reason are strong constraints mainly from CRESST, which is
optimal for inelastic scattering since the heavy tungsten target
nuclei allow for very efficient DM energy loss. In order to illustrate
these constraints we show in fig.~\ref{fig:msigma-plot-inel} allowed
regions in the $\sigma_p, m_\chi$ plane for fixed values of
$\delta$. The upper and lower left panels correspond to the SI fits
with $\delta = 130$~keV and 40~keV, respectively. We observe that the
3$\sigma$ DAMA region at $\delta = 130$~keV is completely excluded by
the CRESST 90\%~CL bound and quite strongly constrained by several
other experiments.
The very thin strip corresponding to the DAMA region in the lower left
plot illustrates the fine tuning problem. Furthermore, we observe that
this solution is marginally compatible with the XENON bound.

\begin{figure}
  \begin{center}
    \begin{tabular}{cc}
      \hspace{.25cm}\includegraphics[width=7.5cm]{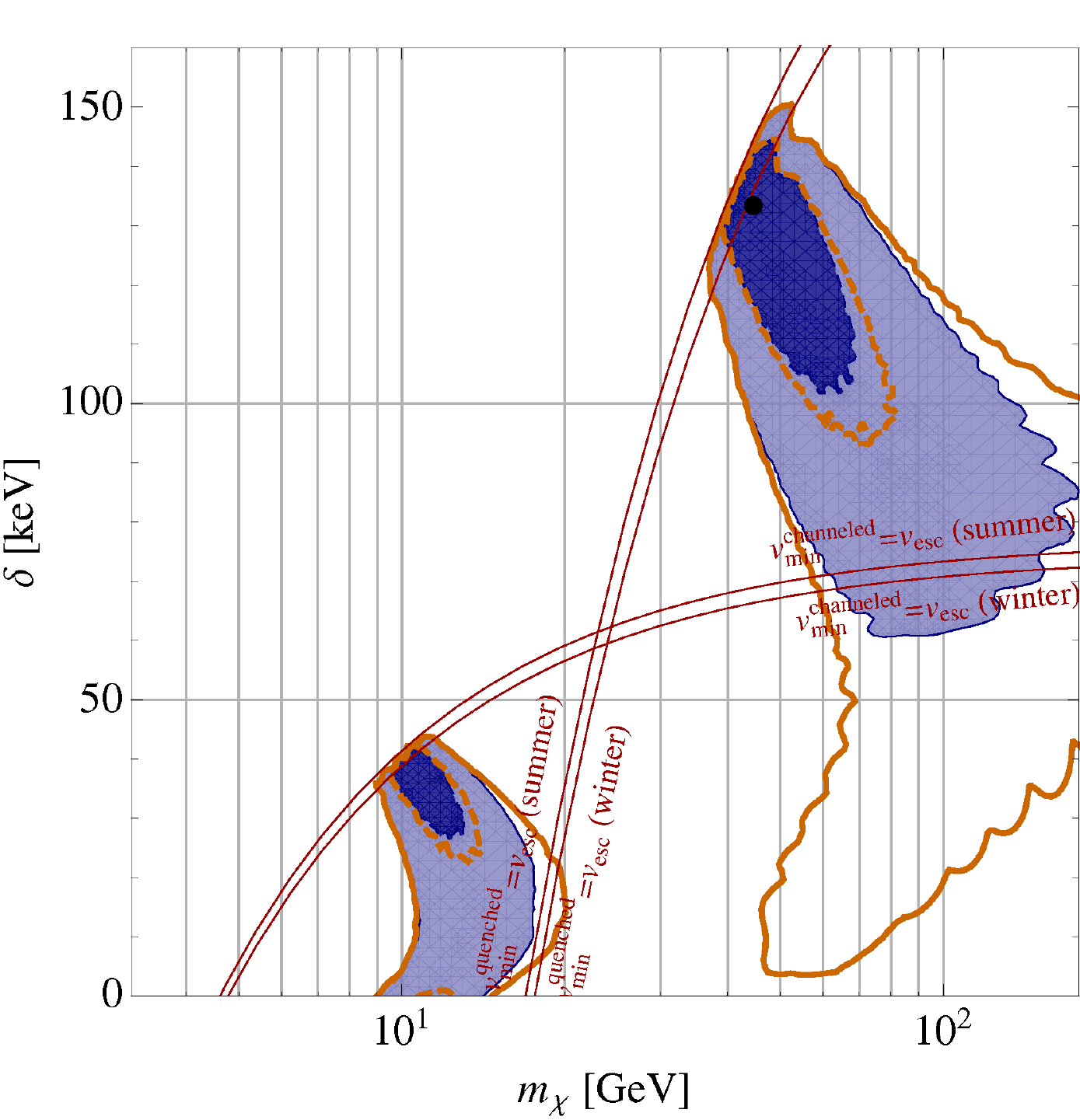} & \\
      \includegraphics[width=8cm]{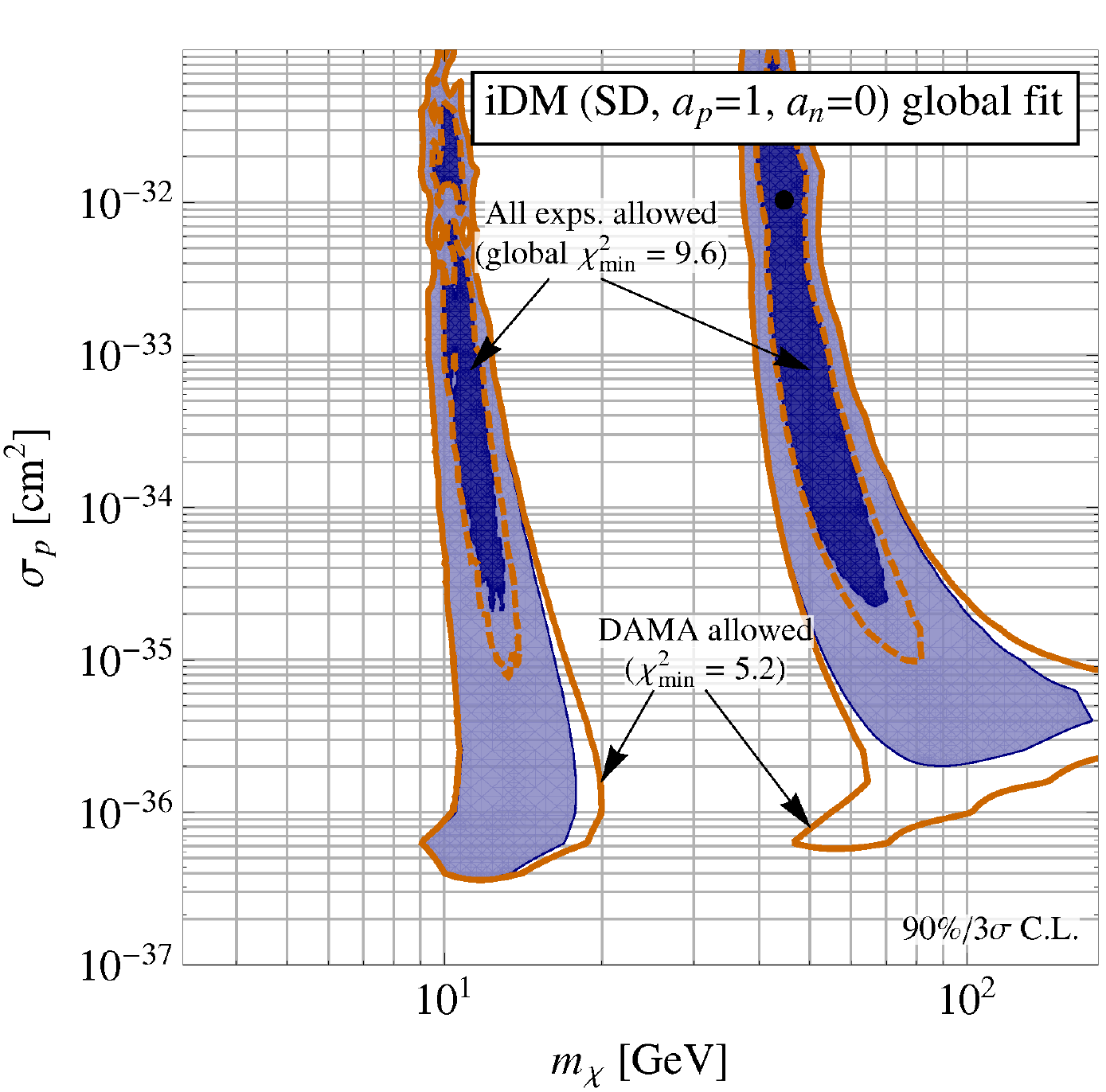} &
      \raisebox{.15cm}{\includegraphics[width=8cm]{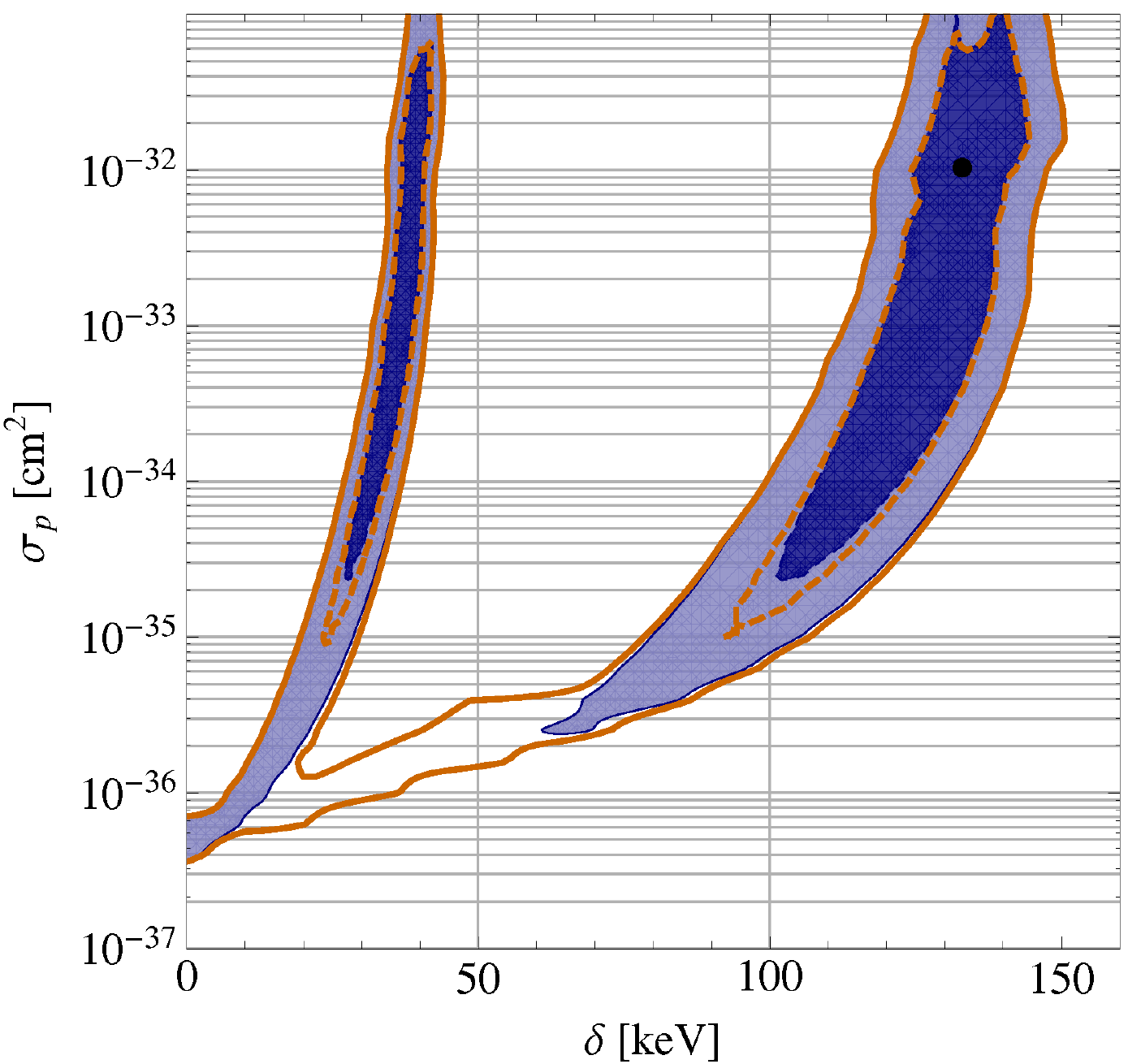}}
    \end{tabular}
  \end{center}
  \caption{Global fit of inelastic, spin-dependent DM. We show
  projections of the 3-dimensional regions at 90\% and 3$\sigma$ CL
  onto the three 2-dimensional planes, by minimizing the global
  $\chi^2$ in each case with respect to the third (un-displayed)
  parameter. Confidence regions are defined for 2~dof. The fit
  includes CDMS (2008 + 2009 data), XENON (2009 analysis), DAMA,
  CRESST-II, ZEPLIN-III, and KIMS. (We omit PICASSO which, due to the
  light target nucleus $^{19}$F has virtually no sensitivity to
  inelastic DM). The shaded regions refer to the global fit, whereas
  the open contours correspond to DAMA data only. In the upper panel
  we show also contours of $v_\mathrm{min} = v_\mathrm{esc} +
  v_\mathrm{earth}$ for a recoil energy of $E = 3$~keV, with and
  without quenching of iodine scatters, where $v_\mathrm{earth}$ is
  the velocity of the earth relative to the halo, depending on the
  time in the year.}
  \label{fig:isd-fit}
\end{figure}

Let us now move to the spin-dependent case. The results of the
parameter scan of the combined $\chi^2$ is shown in
fig.~\ref{fig:isd-fit}.  We observe that the allowed regions are much
larger than in the SI case. There are two different regions
corresponding to fits of similar quality, corresponding to quenched
events in DAMA and channeled events according to
\cite{Bernabei:2007hw}:
\begin{align}
 m_\chi \simeq 40-70 \, {\rm GeV},\quad \delta \simeq 130 \, {\rm keV} &
 \qquad \text{(quenched events)} \\
 m_\chi \simeq 10 \, {\rm GeV},\qquad\quad \delta \simeq 40 \, {\rm keV} &
 \qquad \text{(channeled events)} 
\end{align}
with cross sections in a wide range of ${\rm few} \times 10^{-35} \,
{\rm cm}^2 \lesssim \sigma_p \lesssim {\rm few} \times 10^{-32}\, {\rm
cm}^2$. The right panels of fig.~\ref{fig:msigma-plot-inel} show that
the allowed regions are safely compatible with the constraints from
all other experiments. The explanation is as follows: the SD coupling
to protons drastically reduces the power of even $Z$ target
experiments (XENON, CDMS), while the inelastic kinematics strongly
disfavour light targets (PICASSO), which provide a main challenge for
DAMA in the elastic SD case, see fig.~\ref{fig:msigma-plot-el}.

It is, however, important to remark that for each fixed $\delta$, it
is only a very small range of $m_\chi$ that gives a good fit to the
DAMA data.  Varying $m_\chi$ by a small amount requires a large change
of the cross section to maintain a good fit, see
fig.~\ref{fig:msigma-plot-inel} right panels. This is because the
scattering is sensitive to the exponentially suppressed tail of the DM
velocity distribution close to $v_{\rm esc}$. However, in contrast to
the SI case, we observe from the top panel of fig.~\ref{fig:isd-fit}
the 90\%~CL regions extend relatively far way from the
$v_\mathrm{min}=v_\mathrm{esc}$ curves. Hence, the iSD case does
require a relatively precise tuning of model parameters ($\sigma_p,
m_\chi,\delta$) among themselves, but the tuning with respect to
astrophysics is not necessary here. Therefore, if we assume the
spin-dependent inelastic scenario to be true, the results can be
interpreted as reflecting very high sensitivity of direct detection
experiments to the model parameters up to astrophysical uncertainties.

Let us mention that we have not been able to include data from the
COUPP~\cite{Behnke:2008zza} experiment in our analysis, due to missing
information. In contrast to the elastic SD case we do not
expect that COUPP will provide a relevant constraint in the inelastic
case. Namely, the scattering on $^{19}$F is
negligibe due to its small mass, in the same way as for PICASSO. 
The constraint from iodine contained
in their CF$_{3}$I target should also be much weaker than the one coming
from KIMS due to the much larger exposure of the latter (3409 kg days
for KIMS vs.\ 250 kg days for COUPP). We do not show the results for
iSD scattering off neutrons, but we have checked that for the same
reasons as in the elastic case, the DAMA region is safely excluded by
constraints from XENON and CDMS.

\begin{figure}
  \begin{center}
      \includegraphics[width=0.32\textwidth]{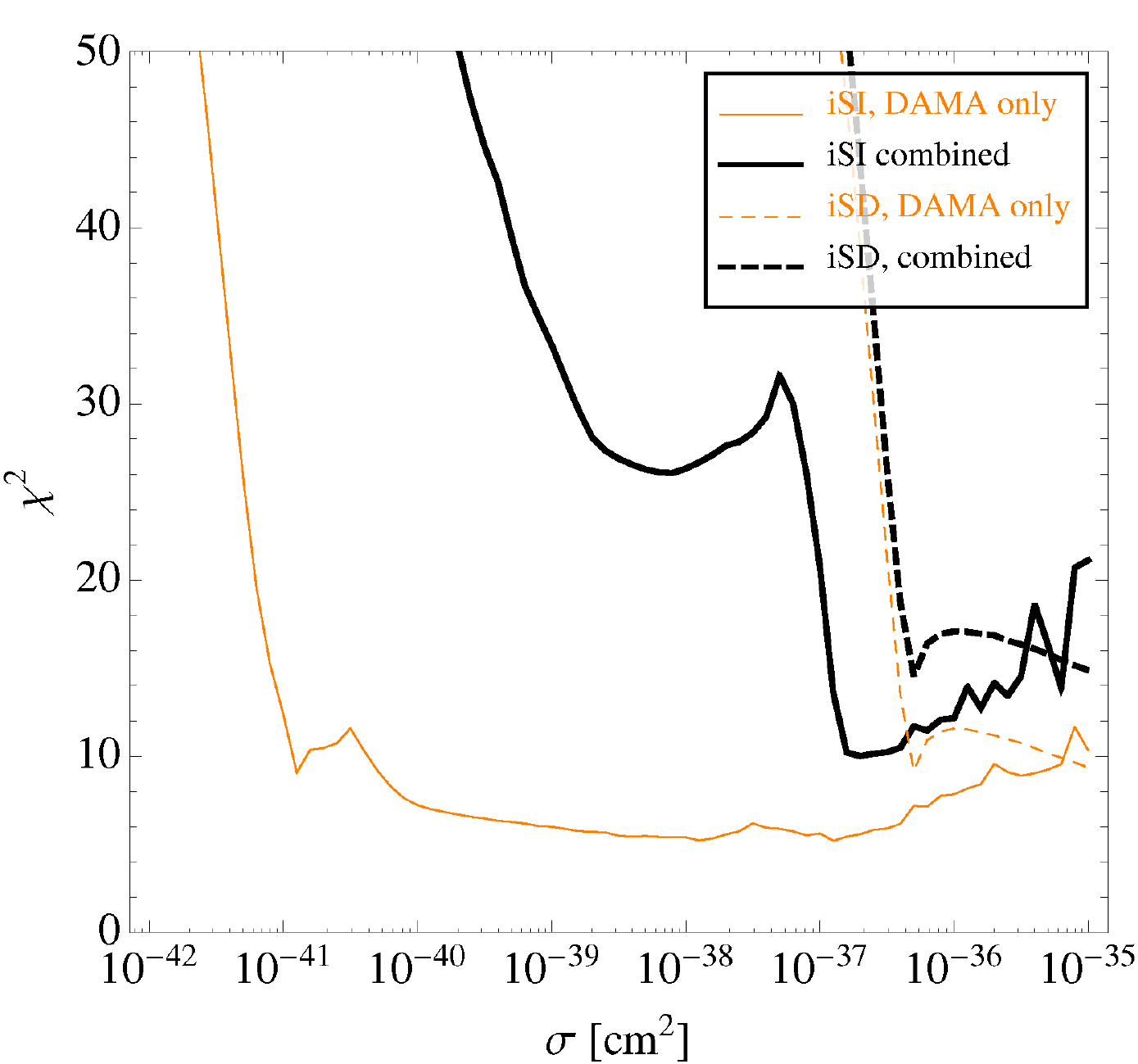}
      \includegraphics[width=0.32\textwidth]{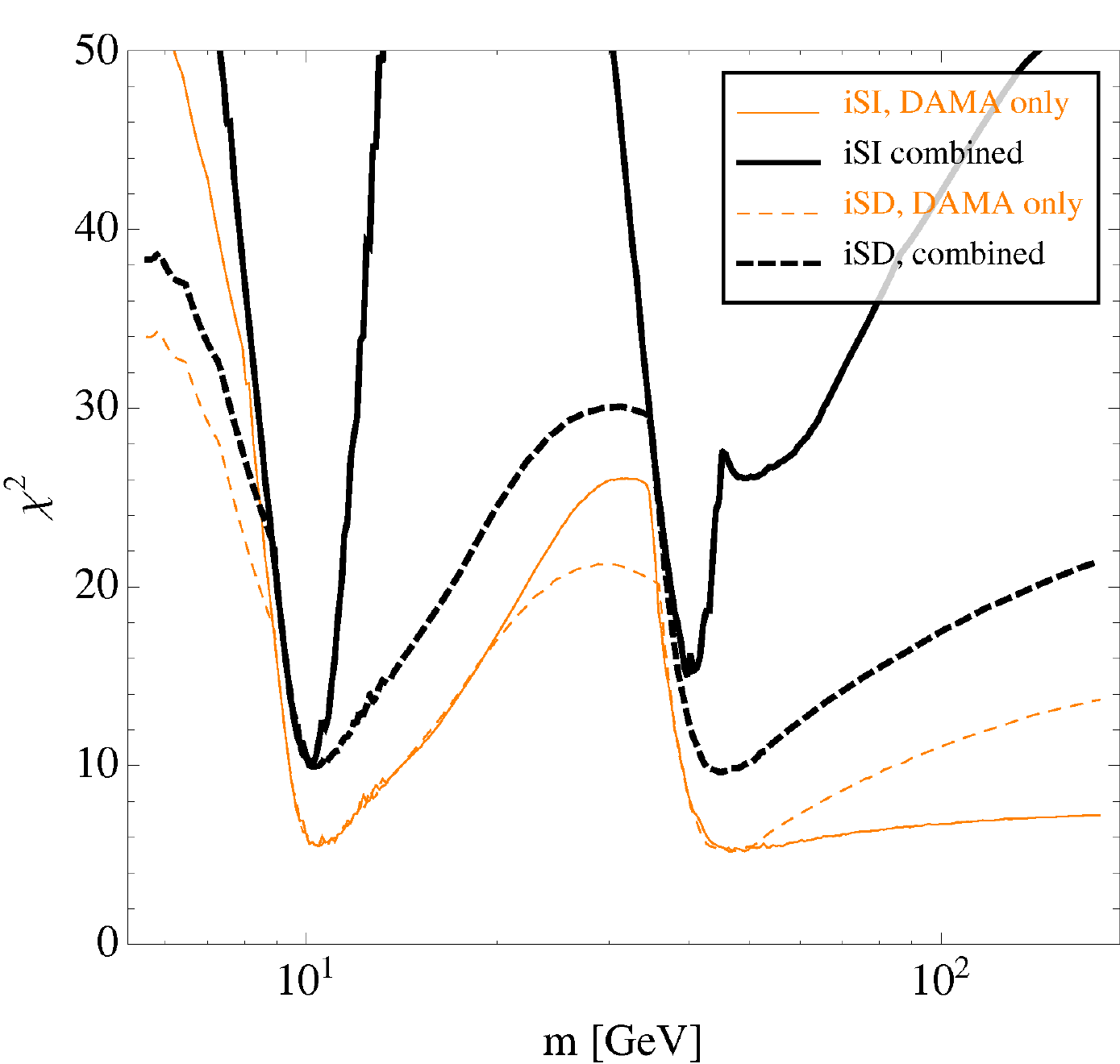}
      \includegraphics[width=0.32\textwidth]{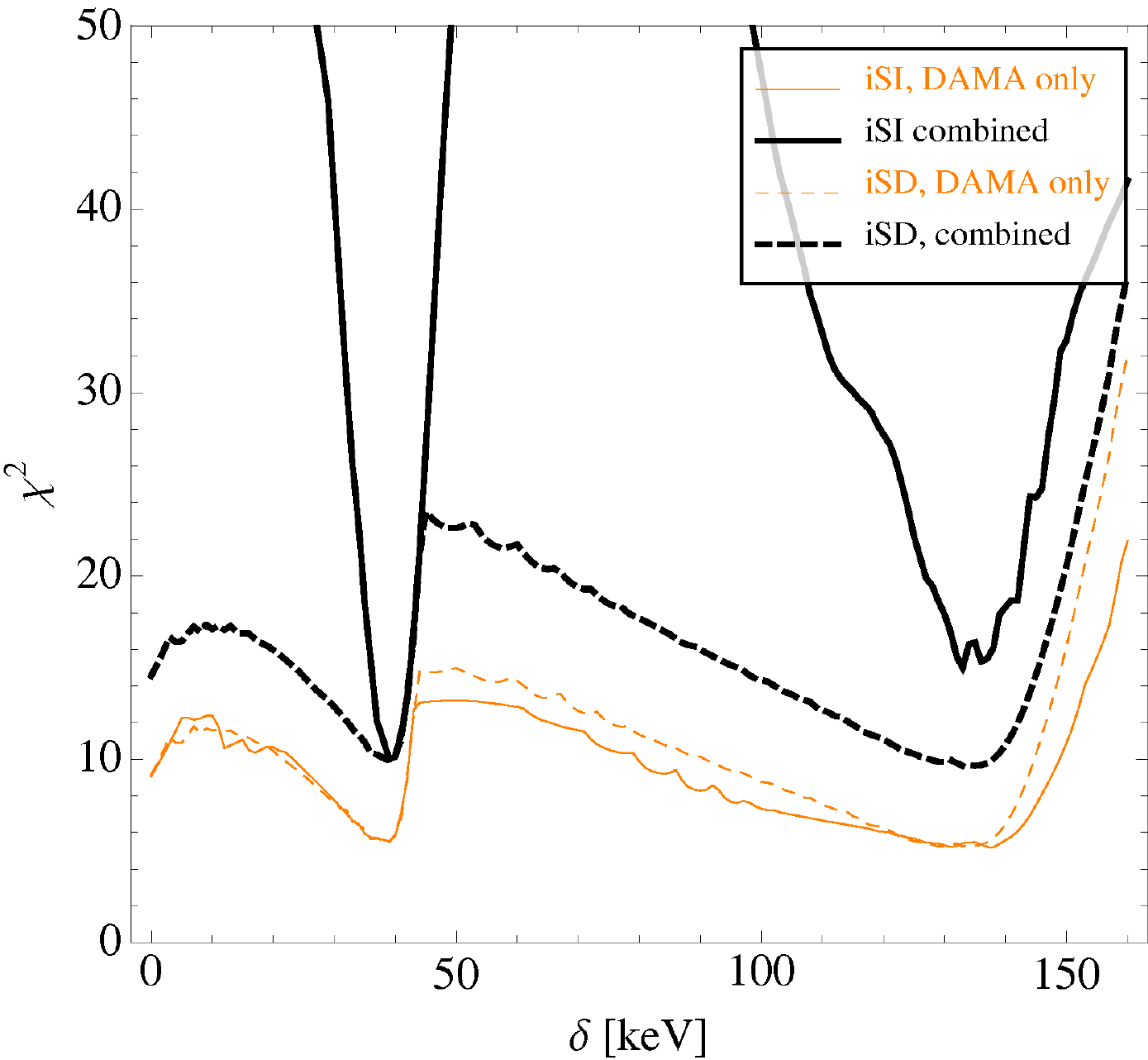}
  \end{center}
  \caption{$\chi^2$ projected onto the $\sigma_p, m_\chi$, and
  $\delta$ axes, where we minimize with respect to the two
  un-displayed parameters. We show the $\chi^2$ for DAMA data only
  (orange), and for the global data (black) assuming iSI interactions
  (solid) and iSD interactions on protons (dashed).}
  \label{fig:chisq}
\end{figure}

To conclude this section, we show in fig.~\ref{fig:chisq} the
projections of the DAMA-only and global $\chi^2$ function for the
three parameters $\sigma_p, m_\chi,\delta$, separately, for the iSI as
well as iSD scenarios.

\section{A simple model}
\label{simple_model}

As we have seen in the previous section, the possibility that DM
scatters inelastically with spin-dependent cross section provides a
viable explanation for the DAMA signal that simultaneously avoids the
bounds from the other direct detection experiments. We now show that
iSD scattering can be realized in concrete models.

It actually does not require too much work to find such a model. To be
concrete let us start with fermionic DM that interacts with the
visible matter through the effective interaction
\beq\label{DM-interaction}
{\cal L}_{\rm int}=\frac{1}{\Lambda^2}\big[\bar \psi \Gamma_{\rm DM}\psi\big]\big[ \bar q \Gamma_{\rm vis}q\big],
\eeq
where $\Gamma_{\rm DM, vis}$ denote the Dirac structure of the
four-fermion operator which we keep general for now, $\psi=(\eta,
\xi^\dagger)$ is a Dirac fermion (for two-component spinors we use the
notation of \cite{Dreiner:2008tw}) and $q$ are the light quark
fields. We could have equally well chosen couplings to leptons, in
which case the analysis of DAMA would have followed
ref.~\cite{Kopp:2009et}. The above effective interaction can arise from an exchange of heavy mediators with mass
${\mathcal O}(\Lambda)$ under which both visible and DM fermions are
charged. In \cite{TuckerSmith:2001hy} the four-fermion interaction
\eqref{DM-interaction} was chosen to be of the $V\otimes V$ form
(i.e., $\Gamma_{\rm DM}=\gamma_\mu, \Gamma_{\rm vis}=\gamma^\mu$).
This gives a spin-independent scattering cross section for inelastic DM. We
show below that tensor interactions give spin-dependent inelastic
scattering instead. To the best of our knowledge this realization has
not been discussed in the context of inelastic DM scattering in the
literature before (for some particle physics realizations of inelastic
DM see \cite{Cui:2009xq, Arina:2009um,ArkaniHamed:2000bq, Alves:2009nf,
Batell:2009vb,Katz:2009qq,Cheung:2009qd, Kadastik:2009gx}).

If $\psi$ is a Dirac fermion, the interaction \eqref{DM-interaction}
leads to elastic scattering. However, if in addition to the Dirac mass
term $m\bar \psi \psi$ there are also Majorana mass terms
$({\delta_\eta}\eta\eta+{\delta_\xi}\xi\xi)/{2}$, then the Dirac
fermion splits into two Majorana fermions with masses $m\pm\delta$
(for simplicity let us take $\delta_\eta=\delta_\xi=\delta$). The mass
eigenstates are \cite{TuckerSmith:2001hy}
\beq
\chi_1=\frac{i}{\sqrt2}(\eta-\xi), \qquad \chi_2=\frac{1}{\sqrt2}(\eta+\xi).
\eeq
It is reasonable to assume that $\delta\ll m$, since Majorana mass
terms break a global symmetry, while the Dirac mass term conserves
it. This is exactly the hierarchy needed phenomenologically, since
$m\sim \co(100~{\rm GeV})$, $\delta \sim \co(100~{\rm keV})$ are
needed for DAMA.

Let us now assume that the four-fermion interaction between DM and
the visible sector \eqref{DM-interaction} is of $T\otimes T$ form,
\beq\label{DM-interaction-tensor}
{\cal L}_{\rm int}=\frac{C_{\rm T}}{\Lambda^2}
\big[\bar \psi \Sigma_{\mu\nu}\psi\big]
\big[ \bar q \Sigma^{\mu\nu}q\big] \,,
\eeq
where $\Sigma^{\mu\nu}=i[\gamma^\mu,\gamma^\nu]/2$. From the relations
$\chi_i\sigma^{\mu\nu}\chi_j= -\chi_j\sigma^{\mu\nu}\chi_i$,
$\chi_i^\dagger\bar \sigma^{\mu\nu}\chi_j^\dagger= -
\chi_j^\dagger\bar \sigma^{\mu\nu}\chi_i^\dagger$
(see e.g.~\cite{Dreiner:2008tw}), it follows that for Majorana fermions the
diagonal tensor operator vanishes.\footnote{Here,
$\sigma^{\mu\nu}\equiv i(\sigma^\mu \bar\sigma^\nu -
\sigma^\nu\bar\sigma^\mu)/4$, $\bar\sigma^{\mu\nu}\equiv
i(\bar\sigma^\mu \sigma^\nu - \bar\sigma^\nu \sigma^\mu)/4$, and
$\sigma^\mu = (1,\sigma^i)$, $\bar\sigma^\mu = (1,-\sigma^i)$, with
$\sigma^i$ the Pauli matrices.} Thus, one finds for the DM tensor
current
\beq \bar \psi
\Sigma_{\mu\nu}\psi = -2i(\chi_2\sigma_{\mu\nu}\chi_1+\chi_2^\dagger\bar
\sigma_{\mu\nu}\chi_1^\dagger) \,,
\eeq 
which leads to inelastic scattering for $\delta \neq 0$. Furthermore,
it is well known that in the nonrelativistic limit the $T\otimes T$
interaction leads to spin dependent scattering, see
e.g.~\cite{Kurylov:2003ra}. For instance, in our case  the matrix
element for $\chi_1\to \chi_2$ scattering on a nucleus $N$ has the
form $\langle N | \bar q S_i q| N\rangle (\zeta_{2,s'}^\dagger
\sigma_i \zeta_{1,s})$,
with $S_i$ the spin operator 
and $\zeta_{2,s'}, \zeta_{1,s}$ the nonrelativistic two-component
spinor wave functions for the DM particles.  

It is important to note that the $T\otimes T$ current is the only chiral
structure that leads to spin-dependent inelastic scattering in the
above simple model of two Majorana fermions split by small Majorana
mass terms. The $T\otimes TA$ structure of the four-fermion
interaction would vanish in the $v\to 0$ limit, while the $TA\otimes
TA$ interaction is equivalent to $T\otimes T$ as is easily checked
from the definition of $\gamma_5$.  The $V \otimes V$ product leads to
SI inelastic scattering \cite{TuckerSmith:2001hy}, $A\otimes A$ to SD
elastic scattering, the $V\otimes A$ product vanishes in the
nonrelativistic limit, while the scalar and pseudoscalar couplings
obviously do not lead to spin dependent interactions.

One still has the freedom to choose appropriate values for Wilson
coefficients $C_T$ for different flavors of the quark current, i.e.\ the
couplings to $u$ and $d$ quarks. In section~\ref{constraints} we
explored two extreme cases where the coupling is proportional to the
charge (so that DM scatters only on protons) or that the coupling to
$d$ is twice as large and opposite to the coupling to $u$ quarks (so
that DM scatters only on neutrons). From the fits the first option is
preferred.

One can also construct a model with DM that is a scalar and that has
inelastic spin-dependent scattering on visible matter. But this
interaction is inevitably suppressed by the nonrelativistic DM
velocity because of the derivative in the DM current and thus less
realistic phenomenologically in light of relatively large scattering
cross sections needed to fit DAMA.

\section{Conclusions}
\label{conclusions}

The CDMS-II collaboration has recently reported the observation of two
events, with an expected background of $0.9$ events \cite{:2009zw}, by
about doubling the exposure with respect to previous results.  We have
performed a combined analysis of direct DM searches including the
claimed signal by DAMA \cite{Bernabei:2003za, Bernabei:2008yi},
the recent CDMS-II results, as well as a re-analysis of XENON10
data. We have considered four classes of possible WIMP-nucleus
scattering models: elastic (e) or inelastic (i) scattering and
spin-dependent (SD) or spin-independent (SI) scattering. This covers a
large set of DM models. While three types of WIMP-nucleus scattering
have already been considered in the literature (i.e., eSD, eSI, and
iSI scattering), our analysis is the first to also include the fourth
possibility --- the inelastic spin dependent scattering. In fact it is
this latter possibility that can simultaneously explain the DAMA
signal and avoid bounds from the other direct detection experiments as
we demonstrated in the present paper.

For eSI scattering the DAMA region is safely excluded by data from
XENON10, based on a recent re-analysis of their
data~\cite{XENON:2009xb}, mainly due to a lower energy threshold,
while for the eSD case DAMA is in conflict with the bound from
PICASSO. For iSI scattering we find two possible solutions. The
traditional ``iDM'' region with $m_\chi \simeq 50$~GeV and $\delta
\simeq 130$~keV is disfavoured by CRESST-II data and further
constrained by CDMS-II. A low mass region exists around $m_\chi \simeq
10$~GeV and $\delta \simeq 40$~keV. It requires, however, the
presence of the channeling effect in DAMA, and leads to rather severe
tuning of the DM mass and mass-splitting with respect to the properties of
the galactic DM halo, such that the WIMP velocity distribution is sampled
precisely around the escape velocity.

For the iSD case, on the other hand, we do find allowed regions of the parameter space such that the signal in 
DAMA is explained, while there is no conflict with any of the other experiments for non-vanishing couplings to
protons and suppressed couplings to neutrons. The solution suffers
from some tuning among the DM parameters $\sigma_p, m_\chi$, and
$\delta$ (which a priori is not a problem, but just indicates very
high sensitivity of the experiments to these parameters), while they do
not require a very precise tuning with respect to the escape velocity.
The required DM parameters for the two solutions are $m_\chi$ around
40--70~GeV, $\delta \simeq 130$~keV (corresponding to quenched events
in DAMA), and $m_\chi \simeq 10$~GeV, $\delta\simeq 40$~keV (which
relies on the channeling effect).  Cross sections are found in a wide
range of ${\rm few} \times 10^{-35} \, {\rm cm}^2 \lesssim \sigma_p
\lesssim {\rm few} \times 10^{-32}\, {\rm cm}^2$.
We presented also a simple toy model, showing how to realize iSD
scattering in a specific framework.

In the case of elastic scattering, we also adopted a somewhat
speculative approach to the recent CDMS-II data, and have performed a
maximum likelihood fit to the two events. We find that at 1$\sigma$~CL
a closed allowed region appears for CDMS (``positive signal''), while
already at 90\%~CL only an upper bound is obtained. The CDMS favoured
region is largely excluded by the XENON bound (and the PICASSO bound,
in case of SD scattering off protons). In a combined analysis of CDMS
and the exclusion limits from other experiments, the ``signal''
becomes less than $1\sigma$. More information on this 
can be expected soon from XENON-100 \cite{Aprile:2009yh}.
\\[8mm]
{\bf Note added in proofs}\\[2mm]
After this work was completed, we were able to perform a fit to the
data from the COUPP experiment. This was made possible by kind
assistance from the COUPP collaboration, especially Juan I. Collar, who
provided crucial data on the bubble formation threshold energies. We
have checked that COUPP limits are subdominant in the case of eSI, iSI,
and iSD scattering. For eSD scattering, COUPP improves the PICASSO limit
by up to a factor of 2 in $\sigma_p$ for $m_\chi \gtrsim 50$~GeV.

\acknowledgments

We would like to thank Jernej Kamenik for useful discussions. 
JK would like to thank the Max-Planck-Institute for Nuclear Physics
for an enjoyable visit and kind hospitality during part of this work.
Fermilab is operated by Fermi Research Alliance, LLC under Contract
No.~DE-AC02-07CH11359 with the US Department of Energy.  This work was
partly supported by the Sonderforschungsbereich TR~27 ``Neutrinos and
Beyond'' of the Deutsche Forschungsgemeinschaft. The work of JZ is supported in part by 
the European Commission RTN network, Contract 
No. MRTN-CT-2006-035482 (FLAVIAnet) and by the 
Slovenian Research Agency. 

\appendix
\section{Interpretation of CoGeNT results}

\begin{figure}
  \begin{center}
      \includegraphics[width=0.7\textwidth]{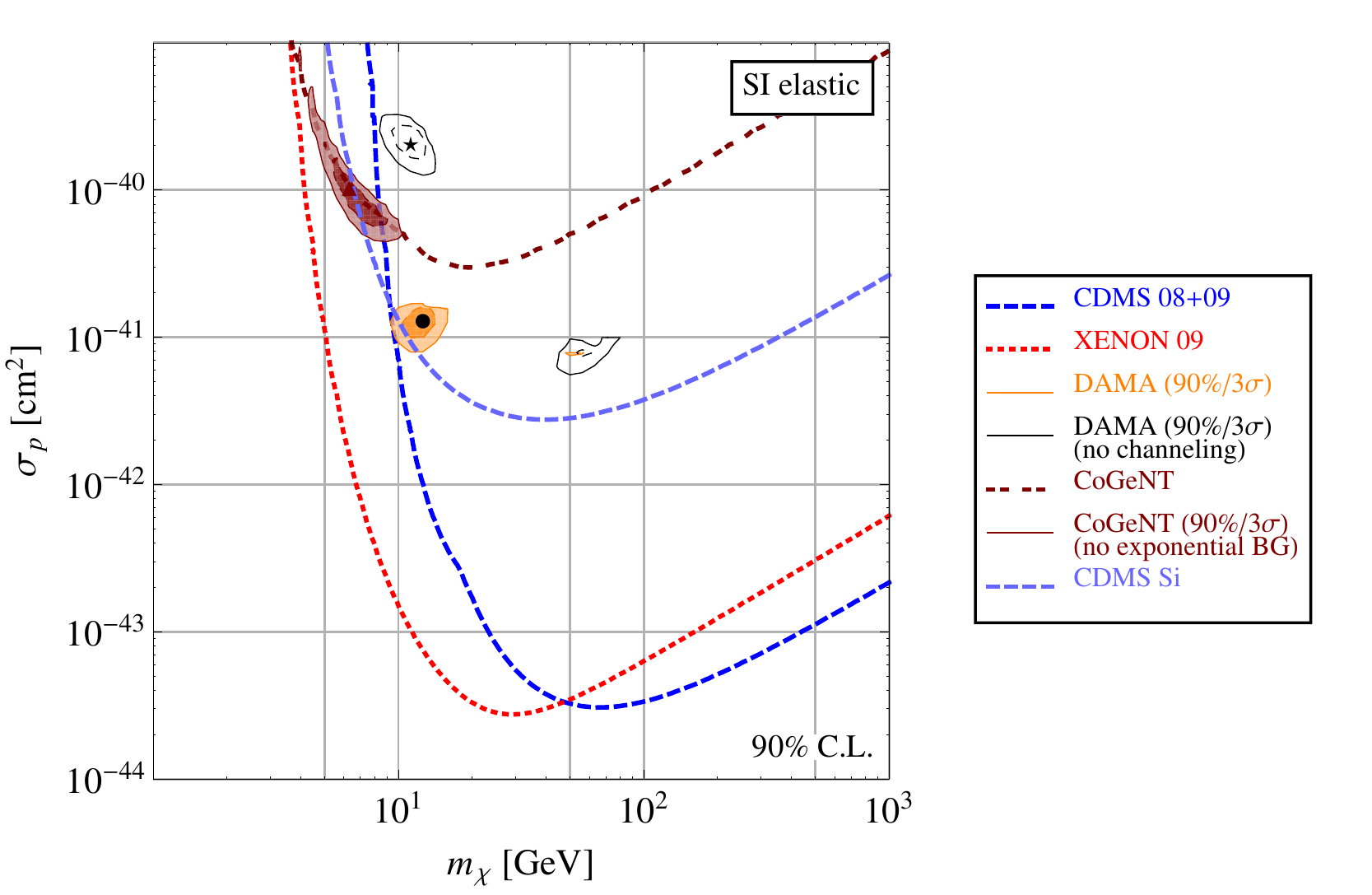}
  \end{center}
  \caption{CoGeNT and DAMA allowed regions (90\% and 3$\sigma$~CL) and constraints
  from other experiments (90\%~CL) for elastic spin-independent scattering. Shaded
  DAMA regions have been obtained assuming channeling according
  to~\cite{Bernabei:2007hw}, while the black contours correspond
  to no channeling. The CoGeNT upper bound follows from a fit including an
  exponential background, while the allowed region is obtained when the exponential
  is removed from the background model.}
  \label{fig:cogent-eSI}
\end{figure}

After publication of this manuscript, the CoGeNT collaboration has published results
from the initial run of their ultra low noise germanium detector in the Soudan
Underground Laboratory~\cite{Aalseth:2010vx}. The detector has a very low
energy threshold of $0.4~{\rm keVee}$, the lowest achieved so far by any dark matter
experiment. The data from an eight week exposure of the detector with a
fiducial mass of $330~{\rm g}$ reveals $\sim 100$ events near the low-energy
threshold for which the collaboration was not able to identify a background
source. The ``signal'' is consistent with an exponential background. It is,
however, also compatible with the hypothesis of an ${\cal O}(10{\rm ~GeV})$ WIMP.
Below we show that this hypothesis is strongly disfavored by other
experiments, though not completely ruled out.

\begin{figure}
  \begin{center}
      \includegraphics[width=\textwidth]{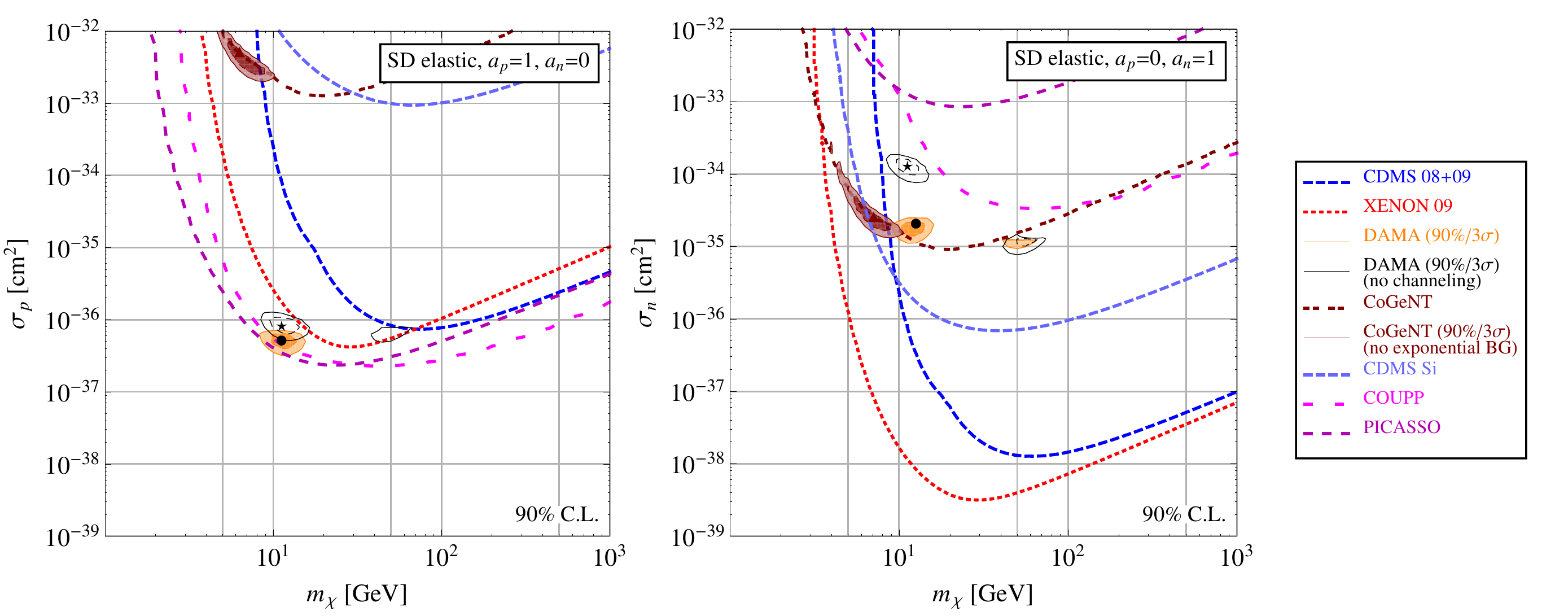}
  \end{center}
  \caption{CoGeNT and DAMA allowed regions (90\% and 3$\sigma$~CL) and constraints
  from other experiments (90\%~CL) for SD scattering on protons (left) and on neutrons (right). See also caption of Fig. \ref{fig:cogent-eSI}.}
  \label{fig:cogent-eSD}
\end{figure}

In the analysis we assume standard astrophysical parameters for the dark matter
halo, see Section~\ref{Event_rates}. In the fit we use the data from Fig.~3 of
\cite{Aalseth:2010vx} in the 0.4--3.2~keVee energy range.  For the quenching
factors we use the approximate formula (based on \cite{Chagani:2008in}, Eq.
(2.6), with $\kappa=0.2$ \cite{Barbeau:2007qi})
\begin{equation}
  \frac{2 }{1 + \sqrt{1 + 15.55/ E}} \,,
\end{equation}
where E is the energy in keVee. Whether or not one finds in the fit a DM signal
depends crucially on the assumed background. We perform fits for two different
background models: (i) a 7-parameter model consisting of a constant (1 parameter), an
exponential (2 parameters), and two Gaussians at the known positions of the
${}^{68}$Ge and ${}^{65}$Zn lines ($2 \times 2$ parameters); this background
model is similar to the one used by the CoGeNT collaboration. (ii) A similar
background model, but without the exponential. The results
are shown in Fig.~\ref{fig:cogent-eSI} for elastic SI scattering, and in
Fig.~\ref{fig:cogent-eSD} for elastic SD scattering. We also include fits
to the data from the CDMS Silicon detectors (labeled
CDMS-Si)~\cite{Akerib:2005kh,Akerib:2005za}. In the SD case, we use the 
structure functions for ${}^{29}$Si as given in~\cite{Bednyakov:2006ux}.

For the full 7-parameter background model only an upper limit on the WIMP
cross-section is obtained for both SI and SD scattering. It is only if one
assumes that the background does not contain an exponential component that a
closed allowed region appears in the $m_\chi-\sigma_p$ plane, with $m_\chi\sim
5-10$~GeV, $\sigma_p\sim10^{-4}$~pb for SI scattering, and
$\sigma_p\sim10^4$~pb ($\sigma_n\sim10$~pb) for SD scattering on protons
(neutrons). In the case of SD scattering on neutrons, the CoGeNT-preferred
region overlaps with the DAMA-preferred region. For SI scattering, marginal
overlap between CoGeNT and DAMA could be achieved if the fraction of channeled
events is assumed to be smaller than what was assumed by the DAMA collaboration
(orange shaded contours), but larger than zero (black contours).  Also, an
admixture of an exponential background to a DM signal in CoGeNT could shift the
allowed region to lower values of $\sigma_p$, closer to the DAMA allowed
region. However, without a clear prediction for the background all such
modifications are mere speculations.  Furthermore, both the DAMA allowed region
and the CoGeNT ``signal'' region are ruled out  at 90\% CL by the  other
experiments for the standard choices of experimental parameters.  XENON-10  and
CDMS-Si rule out the SI case as well as SD scattering on neutrons, while COUPP
and PICASSO rule out SD scattering on protons as an explanation for CoGeNT
and/or DAMA, cf. Figs.  \ref{fig:cogent-eSI} and \ref{fig:cogent-eSD}. As
fig.~\ref{fig:cdmsxenon} shows, relaxing the assumptions on the effective light
yield $L_{\rm eff}$ in XENON-10 can make XENON-10 and CoGeNT (but not DAMA)
marginally compatible at 90\% C.L.  We have also checked that CoGeNT results do
not further constrain inelastic DM (whether SI or SD) discussed in Section
\ref{sec:in}.

In conclusion, we find that the background only hypothesis gives an excellent
fit to the CoGeNT data, while a DM interpretation of this data is disfavored by
other experiments. We differ in this conclusion from the authors of
\cite{Fitzpatrick:2010em}, who find that CoGeNT and DAMA are consistent with
the remaining experiments. This difference may be traced back to the treatment
of $L_{\rm eff}$ for XENON-10. The authors of \cite{Fitzpatrick:2010em} assume
the true $L_{\rm eff}$ to lie at the lower end of the 1-$\sigma$ error bars of
the most conservative measurement available~\cite{Manzur:2009hp}.  Given the
discrepancy between different measurements of $L_{\rm eff}$, the possibility of
such a systematic shift cannot be excluded until systematical errors in the
$L_{\rm eff}$ measurement are better understood. Other differences between our
analysis and that of \cite{Fitzpatrick:2010em} are in the choice of DM halo
parameters and in that the authors of \cite{Fitzpatrick:2010em} demand that the
background does not exceed the DM signal, while we allow both to float freely.

\bibliographystyle{apsrev}
\bibliography{./after_cdms}

\end{document}